\documentclass[prb,twocolumn,showpacs,floatfix,amsmath,amssymb,superscriptaddress] 
{revtex4}
\usepackage[dvips]{graphicx}
\usepackage{bm}
\usepackage{latexsym}
\usepackage{graphicx}
\usepackage{times}
\usepackage{amsmath}
\usepackage{dcolumn}
\usepackage{latexsym,amsmath,amssymb,bm,euscript}
\newcommand{\be}{\begin{equation}}
\newcommand{\ee}{\end{equation}}
\newcommand{\bea}{\begin{eqnarray}}
\newcommand{\eea}{\end{eqnarray}}

\begin{document}

\title{Phase diagram of the frustrated spin ladder}
\author{Toshiya Hikihara}
\affiliation{Department of Physics, Hokkaido University, Sapporo 060-0810, Japan}
\author{Oleg A. Starykh}
\affiliation{Department of Physics and Astronomy, University of Utah, Salt Lake City, UT 84112}
\date{\today}
\begin{abstract}
We re-visit the phase diagram of the frustrated spin-$1/2$ ladder with two competing 
inter-chain antiferromagnetic exchanges, rung coupling $J_\perp$ and diagonal coupling $J_\times$.
We suggest, based on the accurate renormalization group analysis of the low-energy
Hamiltonian of the ladder, that marginal inter-chain current-current interaction plays 
central role in destabilizing previously predicted intermediate columnar dimer phase
in the vicinity of classical degeneracy line $J_\perp = 2J_\times$.
Following this insight we then 
suggest that changing these competing inter-chain exchanges from the previously considered 
antiferromagnetic to the {\sl ferromagnetic} ones 
eliminates the issue of the marginal interactions altogether and
dramatically expands the region of stability of the {\sl columnar dimer phase}.
This analytical prediction is convincingly confirmed by the numerical density matrix renormalization
group and exact diagonalization calculations as well as by
the perturbative calculation in the strong rung-coupling limit.
The phase diagram for ferromagnetic $J_\perp$ and $J_\times$ is determined.

\end{abstract}


\maketitle

\section{Introduction}
\label{section:intro}

Frustrated quantum antiferromagnets have for a long time attracted attention
of both theorists and experimentalists.\cite{Diep05} One of the main reasons for this
continuing focus is the dominant role of quantum fluctuations in stabilizing
various classical and quantum orders in this class of systems. 
Spin ladders represent particularly interesting class of models exhibiting
rich variety of phases \cite{shelton96,nersesyan97,fouet06,meyer09,sheng09}. 
In addition to being interesting in their own right,
spin ladders allow for high precision numerical investigations by 
density matrix renormalization group (DMRG) \cite{schollwoeck05,Hallberg06}
and quantum Monte Carlo techniques. 
One of the most striking features of spin ladders consists in the finding \cite{shelton96}
that generic inter-chain interaction flows toward strong coupling,
resulting in a confinement of gapless spin-1/2 spinon excitations of constituent 
spin chains (which form legs of the ladder) into tightly bound spin-1 pairs (triplons).
Extensive experimental efforts \cite{dagotto99} resulted in observation of one-
and even two-triplons states in La$_4$Sr$_{10}$Cu$_{24}$O$_{41}$ \cite{notbohm07}.
Most recently, an evolution of spin excitations from those of deconfined spinons
at high energies to bound triplet and singlet spinon pairs at low energies has been mapped 
via inelastic neutron scattering in the weakly coupled ladder material CaCu$_2$O$_3$ \cite{lake09}. 

While the standard ladder geometry realizes inter-chain interaction between spins on the legs
in the form of non-frustrated exchange along the rungs of the ladder, a more
complicated geometry is possible as well. In this work we focus on the ladder with {\em frustrated}
inter-chain interactions between leg spins, when the inter-chain exchange takes place
both on rungs (equation \eqref{eq:rung} below) and diagonals (equation \eqref{eq:diag}).
Such geometry is in fact typical for many spin chain materials where the superexchange between
spins proceeds via $90$-degree Cu-O bonds. In particular, well studied spin chain oxides
SrCuO$_2$ and Sr$_2$CuO$_3$ are characterized by the presence of 
(very weak) rung and diagonal exchanges \cite{zaliznyak99} 
which frustrate correlations between chain spins and leads to 
extremely low three-dimensional ordering temperatures.

In this work, we re-visit and resolve one of the outstanding questions
in this field - the appearance of spontaneously dimerized ground state
in frustrated spin ladder model with only pairwise exchange interactions
between microscopic lattice spins. Specifically, the Hamiltonian of the problem reads
\be
H = H_{\rm leg} + H_{\rm rung} + H_{\rm diag}
\label{eq:Ham}
\ee
where
\be
H_{\rm leg} = J \sum_n \Big( {\bf S}_{1,n} \cdot {\bf S}_{1,n+1} + {\bf S}_{2,n} \cdot {\bf S}_{2,n+1} \Big)
\label{eq:leg}
\ee
describes
two isotropic Heisenberg chains with positive (antiferromagnetic) 
nearest-neighbor exchange $J$ while
\be
H_{\rm rung} = J_\perp \sum_n {\bf S}_{1,n} \cdot {\bf S}_{2,n}
\label{eq:rung}
\ee
and
\be
H_{\rm diag} = J_\times \sum_n \Big( {\bf S}_{1,n} \cdot {\bf S}_{2,n+1} +
{\bf S}_{2,n} \cdot {\bf S}_{1,n+1} \Big)
\label{eq:diag}
\ee
describe frustrated inter-chain interaction $H_{\rm inter} = H_{\rm rung} + H_{\rm diag}$.

In the weak-coupling limit ($J_{\perp}, J_\times \ll J$)  
one treats inter-chain interaction
$H_{\rm inter}$  as a perturbation and 
takes continuum limit along the chain direction. It is then easy to see
that for $J_\perp < 2 J_\times$ the ground state is of Haldane type, 
with two spin-1/2 on the rung forming effective spin-1, while for 
$J_\perp > 2 J_\times$ rung pairs form singlets, resulting 
in the rung-singlet (RS) phase.
Transition region $J_\perp \approx 2 J_\times$ between these two well-understood
phases requires careful analysis which is described in Ref.\onlinecite{ladder04}.
It was found there that in the narrow region (boundaries are approximate)
\be
2 J_\times - \frac{5 J_\times^2}{\pi^2 J} < J_\perp < 2 J_\times - \frac{J_\times^2}{\pi^2 J}
\label{range} ,
\ee
the ladder
should exhibit {\sl columnar dimer} (CD) phase in its ground state \cite{ladder04}.
This finding was questioned in several extensive numerical studies 
\cite{hung06,kim08}
which suggested that the CD phase is absent and that instead there is a
direct  transition between Haldane ($J_\perp < 2 J_\times$) and RS 
($J_\perp > 2 J_\times$) ground states.
The most recent work \cite{liu08} on this subject does find the dimerized phase,
although the evidence for this is not particularly strong.

Narrow extent \eqref{range} of the suggested CD order makes 
numerical analysis of the problem difficult. We will argue below that in the 
case of {\em antiferromagnetic} couplings, $J_{\perp}, J_{\times} > 0$, the
situation is even more complicated by the presence of marginally relevant
inter-chain interaction between spin currents (uniform magnetization) of the two chains.
We show that this interaction is responsible for suppressing the CD instability
for not too small inter-chain exchange values ($J_\perp \geq 0.3 J$ or so)
and producing the first order phase transition between the Haldane and RS 
phases instead. We also show that changing the {\em sign} of the inter-chain couplings
to a {\em ferromagnetic} one (so that $J_{\perp}, J_\times  < 0$) effectively
removes the current-current interaction from the problem and allows one
to access the CD phase even for not too small $|J_{\perp,\times}|$ values.
These arguments, derived from renormalization group (RG) analysis described in
Section \ref{section:rg}, are supported by extensive 
DMRG and exact-diagonalization calculations as well as 
the perturbation analysis for the strong rung-coupling limit, 
results of which are reported in
Section \ref{section:FM}. 
The ground-state phase diagram in the $J_{\perp}$ - $J_\times$ plane 
for the ferromagnetic case is also presented there. 
The case of antiferromagnetic couplings between chains
of the ladder is addressed again in Section \ref{section:AF}. 
We conclude by summarizing
our findings in Section \ref{section:discussion}.
Appendix describes an application of our numerical approach to the 
well-understood case of a single frustrated Heisenberg chain
which is known to realize the spontaneously dimerized ground state.

\section{RG analysis}
\label{section:rg}

Low-energy description of the problem is based on the continuum representation
of the spin operator
\be
{\bf S}_n/a_0 = {\bf M}(x) + (-1)^x {\bf N}(x) ,
\ee
where $x = n a_0$ and $a_0$ is the lattice spacing, which we set to unity in what follows.
Uniform ${\bf M}$ and staggered ${\bf N}$ magnetizations represent 
spin fluctuations with momenta near $q=0$ and $\pi/a_0$, correspondingly.
Another very important for the following low-energy degree of freedom is 
staggered dimerization $\epsilon(x)$. It represents fluctuational part of the bond strength
(energy density) between two neighboring spins on the chain,
\be
\epsilon(x) = (-1)^x \Big( {\bf S}_n \cdot {\bf S}_{n+1} - C_{\rm av}\Big).
\ee
The second term, $C_{\rm av}$, represents an {\em average} energy per bond,
which is a position-independent constant. Spontaneously dimerized ground state is
characterized by the finite expectation value of dimerization, $\langle \epsilon\rangle$,
which describes the staggered pattern of strong and weak bonds along the chain.

Low-energy limit of the interchain Hamiltonian is then found to contain
at least 5 couplings that flow under RG. 
\be
H_{\text{inter}}=\int dx \Big(g_1 N + g_2 J + g_3 a^2 M + g_4 E + g_5 K\Big)
\label{H-inter}
\ee
where we introduced the following short-hand notations
\bea
&&N={\bf N_1} \cdot {\bf N_2}, ~J= {\bf M}_{1R} \cdot {\bf M}_{2L} + {\bf M}_{1L} \cdot {\bf M}_{2R}, \nonumber\\
&&M=\partial_x {\bf N_1} \cdot \partial_x {\bf N}_2, ~E=\epsilon_1 \epsilon_2,  \nonumber\\
&&K={\bf M}_{1R} \cdot {\bf M}_{1L} + {\bf M}_{2R} \cdot {\bf M}_{2L}
\eea
The relevant couplings are $g_{1,4}$ which describe coupling between staggered magnetizations
${\bf N}_j$ and staggered dimerizations $\epsilon_j$ of the chains, correspondingly. Marginal couplings
include $g_2$ and $g_5$ which describe current-current interaction between chains ($g_2$)
as well as residual (and naively, marginally irrelevant) in-chain backscattering ($g_5$).
The irrelevant terms contain $g_3$, which is of key importance for generating
(together with $g_2$) the novel $\epsilon_1 \epsilon_2$ term (coupling $g_4$ above). 
This term should be, strictly speaking, be supplemented with 
$g_6 \partial_x \epsilon_1 \partial_x \epsilon_2$ which will certainly be generated (as well as 
other more irrelevant terms) -- but its small initial value, $\sim J_{\perp,\times}^2/J$,
allows us to neglect it.
Initial values of the couplings are:
\bea
&&g_1(0) = J_\perp - 2 J_\times ~,~ g_2(0) = J_\perp + 2 J_\times ~,~ g_3(0) = J_\perp/2 ,\nonumber\\
&& g_4(0) = 0 ~,~ g_5(0) = -0.23 (2\pi v).
\eea
The value of $g_5(0)$ has been estimated in Ref.\ \onlinecite{eggert}.
RG equations for our model \cite{ladder04,kim08} are rather similar to the much studied 
case of the zig-zag ladder \cite{nersesyan98}. They are conveniently formulated
in terms of dimensionless variables
\be
G_{1,3,4}=\frac{g_{1,3,4} \lambda^2}{2\pi v}, ~G_{2,5}=\frac{g_{2,5}}{2\pi v}
\ee
and read ($\dot{g} = dg/d\ell$, where $\ell$ is the RG scale)
\bea
\dot{G}_1 &=& G_1 + G_1 G_2 - \frac{1}{2} G_1 G_5 + G_2G_3 - \frac{1}{2} G_2 G_4,\nonumber\\
\dot{G}_2 &=&G_2^2 + 4G_1^2 + 4 G_1 G_3 - 4 G_1 G_4 - 2 G_3 G_4,\nonumber\\
\dot{G}_3 &=&-G_3,\nonumber\\
\dot{G}_4 &=&G_4 - \frac{3}{2} G_1 G_2 - \frac{3}{2} G_2 G_3 + \frac{3}{2} G_4 G_5,\nonumber\\
\dot{G}_5 &=&G_5^2 - 2 G_1^2 + 2 G_1 G_3 + 2G_4^2 .
\label{rg-full}
\eea

Almost everywhere in parameter space $(J_\perp, J_\times)$ $g_{1,4}$ diverge 
exponentially while $g_3$
dies off exponentially and the flow is completely determined by the initial value 
of $g_{1}(0)$ (remember that $g_4(0)=0$).
But in the vicinity of $J_\perp=2J_\times$ line things are different because 
naively there $g_1(0)=0$ as well.
Thus there two relevant operators appear to be absent due to fine-tuning. 
This region requires careful study.

Let us assume that we are very close to this line so that $G_{1}(0) = O(J_\times^2/J^2)$. 
Note that other couplings are $G_{2,3}(0) = O(J_\times/J)$ because they are not sensitive 
to the difference $J_\perp - 2 J_\times$.
In order to understand how the flow starts consider very short RG times, $\ell \leq \ell_0 = 1$
and keep only terms of order $(J_\times/J)^2$ in the right-hand-side of the RG equations. 
The system simplifies dramatically and we find 
that $G_{2,3,5}$ `decouple' from the equations for the relevant $G_{1,4}$ couplings
\bea
G_2(\ell) &=& G_2(0)/(1 - G_2(0)\ell) ,\nonumber\\
G_3(\ell)&=&G_3(0) e^{-\ell} ,\nonumber\\
G_5(\ell) &=& G_5(0)/(1 + |G_5(0)|\ell) \rightarrow -1/\ell .
\label{rg-G2}
\eea
Thus for $\ell \sim 1$ we have
\bea
\dot{G}_1&=&G_1 + G_2 G_3 ,\nonumber\\
\dot{G}_4&=&G_4 - \frac{3}{2} G_2 G_3 .
\label{rg-short}
\eea
Note that for such short $\ell$'s we can safely treat marginal couplings $G_{2,5}$ as constants,
because $G_{2,5}(0) \ll 1$.
The last two equations then admit straightforward solution
\bea
G_1(\ell) &=& (G_1(0) + \frac{q}{2}) e^\ell -  \frac{q}{2} e^{-\ell} ,\nonumber\\
G_4(\ell) &=& \frac{-3q}{2} \sinh [\ell] ,
\label{rg-short-solution}
\eea
where $q\equiv G_2(0) G_3(0)=(2J_\times/2\pi v)^2 = (2 J_\times/\pi^2 J)^2$, 
and we made use of the initial condition $G_4(0) = 0$ and of the fact that the spin velocity
is given by $v=\pi J/2$.
We observe renormalization of the initial values of $G_{1,4}$ couplings by quantum fluctuations.
It is clear that for intermediate range of $0 < \ell \ll \ell_1 = \ln(J/J_\times)$, where the approximation
\eqref{rg-short} is expected to work, the two
equations \eqref{rg-short-solution} describe evolution of relevant $G_{1,4}$ couplings 
with initial values given by $G_1(0) \to G_1(0) + q/2$ and $G_4(0) \to -3q/4$, correspondingly.
The same renormalization can be obtained via direct perturbative calculation,
as was done originally in Ref.\ \onlinecite{ladder04}.

Once $G_{1,4}$ got some non-zero initial values of the order $(J_\times/J)^2$, they will grow
exponentially fast and at $\ell_1 \sim \ln(J/J_\times) \ll \ell_2 =1/G_2(0) \sim J/J_\times$ 
reach values of the order $J_\times/J$.
Note that $G_2(\ell_1) \sim J_\times/J \ll 1$ at this scale and our weak-coupling 
consideration still makes sense. An estimate of the range where the dimerized phase is
expected to appear, {\em neglecting} the flow of the marginal couplings, can now be obtained
 with the help of refermionization procedure as described in Ref.\ \onlinecite{ladder04}.
Specifically, the model \eqref{H-inter} with two competing relevant couplings $g_1$ and $g_4$
maps onto the theory of four (weakly interacting) Majorana fermions which are 
organized into a {\em triplet} with mass $m_t \propto 2\pi v (G_1 - G_4)$ 
and a {\em singlet} with mass $m_s \propto -2\pi v (3G_1 + G_4)$.
The Haldane-to-CD transition corresponds to the vanishing of the triplet mass, $m_t=0$,
and takes place when $J_\perp - 2 J_\times + 5 J_\times^2/(\pi^2 J) =0$.
The CD-to-RS transition is described by the condition $m_s=0$ and results in 
$J_\perp - 2 J_\times + J_\times^2/(\pi^2 J) = 0$. Putting these two boundaries together leads to
the estimate \eqref{range}.

The neglect of the marginal terms, and in particular of the inter-chain
current-current one, $G_2$, is well justified in the asymptotic limit $J_{\perp,\times}/J \to 0$. 
Away from this limit, but still keeping $J_{\perp,\times}/J \ll 1$, one has to be
mindful of the marginally relevant character of $G_2$ term in the case of
{\em antiferromagnetic} (positive) sign of exchanges $J_{\perp,\times}$.
Left alone, $G_2$ would diverge at $\ell \sim J/J_\times$, as follows from 
\eqref{rg-G2}.
It may happen that for not too small $J_{\perp,\times}/J$ the marginal
coupling $G_2$ reaches value of the order 1 before the relevant couplings $G_{1,4}$ do so.
In the following we denote the corresponding RG scale as $\ell_3$: 
thus $G_2(\ell_3) \approx 1$ while $G_{1,4}(\ell_3) < 1$.

Such a behavior implies direct first-order transition between Haldane and RS 
phases \cite{allen00,kim00,tsvelik03,ladder04}, which can be
analyzed with the help of abelian bosonization along the lines sketched previously in Ref.\onlinecite{ladder04}.
In terms of conjugate bosonic fields $\varphi_m$ and $\theta_m$, where $m=\pm$ denotes 
symmetric/antisymmetric combination of the original chain fields,
the interchain interaction \eqref{H-inter} acquires the form
\bea
H'_{\rm inter}& = &\frac{1}{2\pi^2 a_0^2} \int dx \Big( g_2 \cos[\sqrt{4\pi} \varphi_+]\cos[\sqrt{4\pi} \theta_{-}] +
\nonumber\\
&& 2 g_1 \cos[\sqrt{4\pi} \theta_{-}] - (g_1 - g_4) \cos[\sqrt{4\pi} \varphi_+]  
\label{Hp-inter}\\
&& +(g_1 + g_4) \cos[\sqrt{4\pi} \varphi_{-}]\Big). \nonumber 
\eea
This expression includes only relevant and marginally relevant terms of \eqref{H-inter}
(i.e. terms N, E and J) {\sl evaluated} at $\ell = \ell_3$. The two remaining terms
(M and K) are omitted because of their irrelevancy at this late stage of RG development.
In the regime we are interested here $g_2 \gg g_{1,4}$ which implies that
$H'_{\rm inter}$ is minimized by (a) $\varphi_{+} = \sqrt{\pi}/2, \theta_{-} =0$,
and (b) $\varphi_{+} = 0, \theta_{-} =\sqrt{\pi}/2$. The first choice describes
Haldane state while the second corresponds to the RS one, see Ref.\onlinecite{ladder04}.
Since $\theta_{-}$ is pinned in both states, the last term in \eqref{Hp-inter} effectively averages
to zero, $\cos[\sqrt{4\pi} \varphi_{-}] \to 0$. Treating the remaining two terms as a
perturbation, we find that $g_4 = 3g_1$ describes a {\sl first-order transition} line
separating Haldane and RS phases. Everywhere on this line the energies of the
two states are equal. Line's endpoints can be estimated as points where $g_2 = g_{1,4}$, 
see Ref.\onlinecite{ladder04} and Figure~1 therein. Since the Haldane and the RS
states are degenerate on the first-order transition line, elementary excitations on this
line are (massless) domain walls or kinks, interpolating between vacua of types (a) and (b) above.
The spin of such a kink can be evaluated as 
\be
S^z_{\rm total} = S^z_1 + S^z_2 = \partial_x (\varphi_1 + \varphi_2)/\sqrt{2\pi} = \partial_x \varphi_{+}/\sqrt{\pi}.
\ee
Assuming, for example,  $\varphi_{+}(x=-\infty) = 0$ and $\varphi_{+}(x= +\infty) = \sqrt{\pi}/2$, we
find $S^z_{\rm total} = 1/2$: the kink is a {\sl spinon}! \cite{tsvelik03} These spinons can be visualized
as spin-1/2 end states of the Haldane phase segments -- since the length of the segment
is arbitrary, the spinons are mobile and massless \cite{ladder04}. Away from the line
$g_4 = 3 g_1$ the spinons are bound into pairs as the energy cost of two distant kinks
is proportional to the length of `minority' phase segment between them.

Once the initial parameters are such
that one of the relevant couplings $g_{1,4}$ reaches $1$ before the marginal $g_2$, the
intermediate phase separating Haldane and RS phases becomes unavoidable.
For positive $g_4$ such intermediate phase realizes {\sl staggered dimer} (SD) phase
($\varphi_{+} = \sqrt{\pi}/2, \varphi_{-} =  \sqrt{\pi}/2$) while {\sl negative} $g_4$ results
in the CD phase ($\varphi_{+} = 0, \varphi_{-} =  0$) which is the focus
of this work. Such a situation is realized in the strict weak-coupling limit, 
$J_{\perp,\times}/J \to 0$, when the relevant couplings are certain to flow to values
of order one well before any of the marginal ones can do so -- and thus we once
again conclude that dimerized phase is unavoidable. Its numerical detection is
however very challenging in this very limit.

Solving full system of RG equations \eqref{rg-full} numerically we find that 
conditions favoring the direct first order transition are 
realized for (approximately) $J_\times \geq J_\times^* \approx 0.3$ and $J_\perp \approx 2 J_\times$.
Even though it is not {\em a priori} clear that our weak-coupling theory can adequately
describe these somewhat large inter-chain couplings, observation of the direct transition 
between Haldane and RS states in several numerical studies suggests
that analytical description based on \eqref{rg-full} and \eqref{Hp-inter} remains valid there.
We thus conclude that antiferromagnetically coupled frustrated ladder
should exhibit CD phase in the range approximately given by \eqref{range}
and for {\em not too large} inter-chain exchanges, $J_{\times} \leq J_\times^*$.  
Stronger inter-chain exchange leads to the collapse of the CD phase and
direct first-order transition between Haldane and RS ground states.
The situation is
very similar to the phase diagram of the extended Hubbard model,\cite{Nakamura1999,Nakamura2000,furusaki02}
where one finds that charge-density-wave and spin-density-wave phases are
separated by the bond-charge-density-wave phase. This intermediate phase
collapses and gets replaced by the line of direct first-order transition once
Hubbard repulsion constant exceeds some critical value. 

\begin{figure}[t]
\begin{center}
\includegraphics[width=0.45\textwidth]{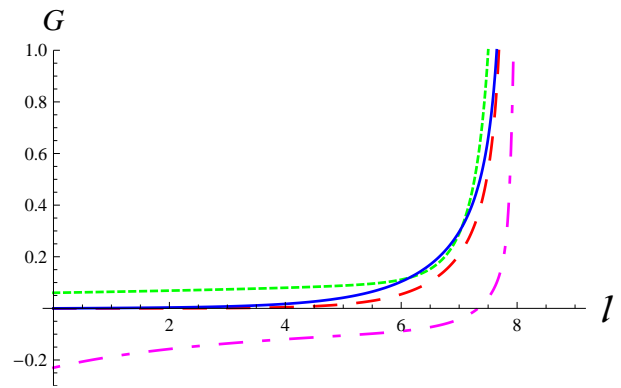}
\end{center}
\caption{
(Color online) RG flow of couplings $G_{1,2,4,5}$ for the case of antiferromagnetic 
inter-chain exchanges $J_\times = 0.15, J_\perp = 0.296$. Notations are as follows:
$G_1$ (red/dashed), $G_2$ (green/dotted), $-G_4$ (blue/solid), $G_5$ (magenta/dot-dashed).
$\ell$ denotes RG scale.}
\label{fig:+015}
\end{figure}

\begin{figure}[t]
\begin{center}
\includegraphics[width=0.45\textwidth]{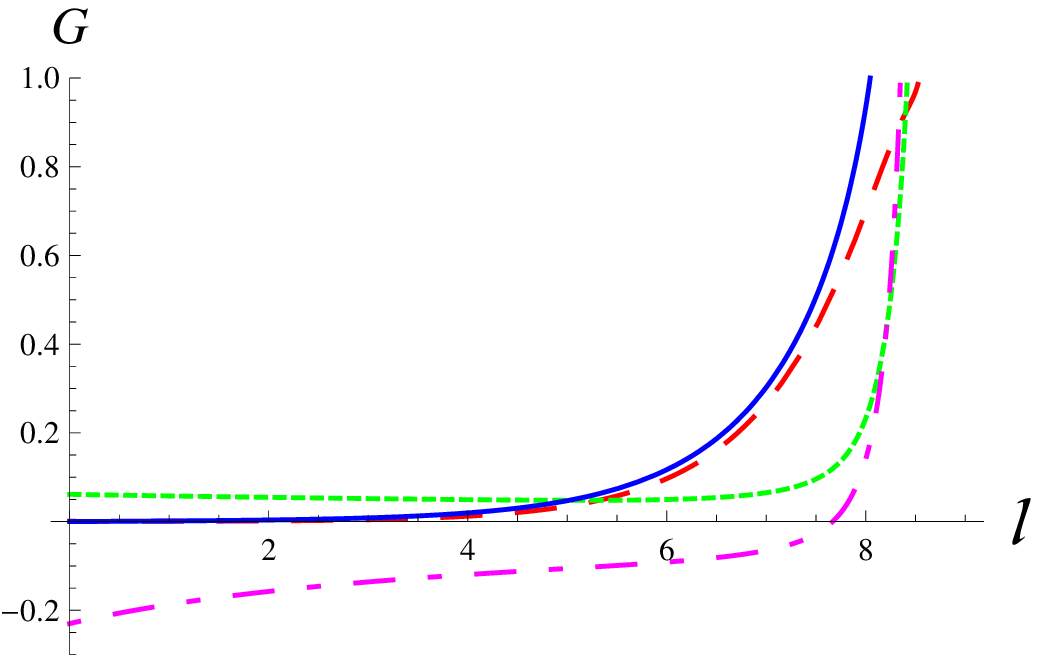}
\end{center}
\caption{
(Color online) RG flow of couplings $G_{1,2,4,5}$ for the case of ferromagnetic 
inter-chain exchanges $J_\times = -0.15, J_\perp = - 0.306$. Notations are as follows:
$-G_1$ (red/dashed), $-G_2$ (green/dotted), $-G_4$ (blue/solid), $G_5$ (magenta/dot-dashed).
$\ell$ denotes RG scale.}
\label{fig:-015}
\end{figure}

The outlined reasoning suggests simple way to avoid the marginal interaction issue altogether: 
all we need to do is to change the signs of both inter-chain exchanges to the {\sl ferromagnetic} (negative) ones.
This simple change {\em preserves} frustrated nature of the inter-chain couplings 
-- large negative $J_\perp$ leads to an effective spin-1 chain and the Haldane phase while
large negative $J_\times$ forces rung spins into the RS phase.
Importantly, this change makes inter-chain $G_2$ 
{\em marginally irrelevant} so that $G_2(\ell) = G_2(0)/(1 + |G_2(0)|\ell) \to -1/\ell$
(neglecting for a moment effect of all other couplings). In effect, this simple sign change
sends the scale $\ell_3$ to infinity, $\ell_3 \to \infty$.
We now should be able, within the weak-coupling approximation,
to get rid of $G_2$ which forces the CD phase
to shrink, without suppressing the all-important competition between $G_1$ and $G_4$ terms.
This is indeed observed in numerical solution of the system \eqref{rg-full}. The difference
between antiferromagnetic and ferromagnetic inter-chain couplings is illustrated in 
Figs. \ref{fig:+015} and \ref{fig:-015}. Coupling $G_2$, shown by dotted  (green) line there, overtakes
relevant $G_{1,4}$ on the way to strong coupling in Fig. \ref{fig:+015} (antiferromagnetic case)
while it remains small in Fig. \ref{fig:-015} (ferromagnetic case).

Results of numerical solutions for some sample values of $J_{\times,\perp}$ are
summarized in the Table~\ref{table1}, which shows that making the inter-chain
exchanges ferromagnetic indeed helps one to unmask the novel columnar dimer
phase.

\begin{table}
\caption{\label{table1} Brief summary of numerical solution of RG system \eqref{rg-full}
for several values of $J_\times$. Positive (negative) values of $J_{\times,\perp}$ correspond
to antiferromagnetic (ferromagnetic) inter-chain exchanges. Second (third) column describes
the range of $J_\perp$ for which the columnar dimer phase is realized according to Eq.\eqref{rg-full}
(Eq.\eqref{range}). Last column shows the most divergent coupling constant which reaches value of order one first.}
\begin{ruledtabular}
\begin{tabular}{cccc} 
$J_\times$ &  range of $J_\perp$ (eq.\eqref{rg-full})    &   estimate of $J_\perp$ (eq.\eqref{range})  
&  leading g  \\
\hline
0.05   & (0.09945, 0.09955) & (0.0987, 0.0997) & $g_4$  \\
0.1   & (0.1978, 0.1982) & (0.1949, 0.1990) & $g_4$\\
0.15   &  (0.295, 0.2956) & (0.2886, 0.2977) &  $g_4$ \\
0.2   & (0.39096, 0.392) & (0.3797, 0.3959) &  $g_4$ \\
0.3   & none & (0.5544, 0.5909) &  $g_2 $ \\
0.4   & none & (0.7189, 0.7838) &   $g_2 $\\
0.5 & none & (0.8733, 0.9747) & $g_2 $\\
	\hline
-0.15 & (-0.303, -0.3062) & (-0.3023, -0.3114)& $g_4$\\
-0.3 & (-0.608, -0.635) & (- 0.6091, -0.6456) & $g_4$\\
-0.4 & (-0.815, -0.882) & (-0.8162, -0.8811) & $g_4$\\
-0.5 & (-1.02, -1.16) & (-1.025, -1.127) & $g_4$\\
\end{tabular}
\end{ruledtabular}
\end{table}

\section{Ferromagnetic Inter-chain couplings}
\label{section:FM}
Motivated by the result of RG analysis in Sec.\ \ref{section:rg}, 
we study the case of ferromagnetic inter-chain couplings 
$J_{\perp}, J_\times <0$.
First, we treat the limit of strong rung coupling, 
$|J_\perp| \gg J,  |J_\times|$, and show that for 
$J + J_\times = 0$ the model exhibits the CD long-range order.
We then present our numerical DMRG and exact-diagonalization 
data for the model.
Combining the results, we determine the ground-state phase diagram, 
which includes the CD phase in a wide parameter range 
between the Haldane and RS phases.

\subsection{Strong rung-coupling limit}
\label{subsection:perturb}

We consider the limit of strong rung coupling, 
$|J_\perp| \gg J, |J_\times|$.
We first diagonalize the rung Hamiltonian $H_{\rm rung}$, 
whose ground states are a direct product of triplet states in each rung.
We then include the effect of 
$H_{\rm leg}$ and $H_{\rm diag}$ perturbatively.
It is convenient to rewrite the perturbation term as
\begin{eqnarray}
H' &=& H_{\rm leg} + H_{\rm diag}
\nonumber \\
&=& \frac{1}{2}(J + J_\times) 
\sum_n ({\bf S}_{1,n} + {\bf S}_{2,n}) \cdot 
       ({\bf S}_{1,n+1} + {\bf S}_{2,n+1})
\nonumber \\
&&+ \frac{1}{2}(J - J_\times) 
\sum_n ({\bf S}_{1,n} - {\bf S}_{2,n}) \cdot 
       ({\bf S}_{1,n+1} - {\bf S}_{2,n+1}).
\nonumber \\
\label{eq:Ham_perturb}
\end{eqnarray}
Note that the first term preserves the total spin in each rung 
while the second term changes the rung-triplet state 
to rung-singlet one and vice versa.

When $J + J_\times \ne 0$, the calculation is easy.
The first term in Eq.\ (\ref{eq:Ham_perturb}) gives 
a nonzero contribution at the first order perturbation 
and lifts the ground state degeneracy of $H_{\rm rung}$.
The effective Hamiltonian turns out to be the spin-1 Heisenberg chain,
\begin{eqnarray}
\tilde{H}^{(1)} 
= \tilde{J}^{(1)} \sum_n \tilde{\bf S}_n \cdot \tilde{\bf S}_{n+1},
\end{eqnarray}
where $\tilde{\bf S}_n$ is the spin-1 operator consisting 
of rung spins ${\bf S}_{1,n}$ and ${\bf S}_{2,n}$ 
and $\tilde{J}^{(1)}=(J + J_\times)/2$.
Therefore, if $J + J_\times > 0$, the system is in the Haldane phase, 
while the system exhibits the ferromagnetic ground state for 
$J + J_\times < 0$.

For $J + J_\times = 0$, the first-order perturbation vanishes, 
and we must turn to the second order.
From a straightforward calculation, we obtain 
the second-order perturbation Hamiltonian of the form,
\begin{eqnarray}
\tilde{H}^{(2)} 
= \tilde{J}^{(2)} \sum_n \left[
 \left( \tilde{\bf S}_n \cdot \tilde{\bf S}_{n+1} \right)^2 - 1 \right],
\end{eqnarray}
with
\begin{eqnarray}
\tilde{J}^{(2)} = - \frac{1}{8 |J_\perp|} (J-J_\times)^2
= - \frac{J^2}{2 |J_\perp|} = - \frac{J_\times^2}{2 |J_\perp|}.
\end{eqnarray}
Therefore, the low-energy physics of the system is described 
by the spin-1 pure biquadratic chain with negative $\tilde{J}^{(2)}$.
For this case it has been established that the model has 
the dimerized ground state.\cite{Parkinson1987,Parkinson1988,BarberB1989,Klumper1989,Affleck1990,Xian1993}
Hence, mapping the spin-1 dimerized phase back to our model,
we conclude that the spin-1/2 two-leg frustrated ladder 
(\ref{eq:Ham}) must exhibit CD phase 
along the line $J_\times = -J$ in the strong rung-exchange limit.

\subsection{DMRG results}
\label{subsection:dmrg}

To search for the CD state and 
determine the ground-state phase diagram, 
we carry out the DMRG calculation\cite{White1992,White1993}
for the frustrated ladder (\ref{eq:Ham}).
The calculation is performed for the system with up to $L=192$ rungs.
For the efficiency of the DMRG method, 
the open boundary condition is imposed in the calculation.
The number of kept states are typically $m=350$ for 
$L \le 96$, $m=400$ for $L=128, 192$, and up to $m=500$ for some cases of 
the severe truncation error.
We have monitored the truncation error of the data by comparing 
the results obtained with different $m$'s 
and confirmed that the $m$ convergence has been achieved 
for the data shown in the following.

To detect the CD order, we calculate the local CD operator 
in the open ladder with $L$ rungs,
\begin{eqnarray}
D_{\rm CD}(n; L) = \sum_{j=1,2} \left(
\langle {\bf S}_{j,n} \cdot {\bf S}_{j,n+1} \rangle
- \langle {\bf S}_{j,n+1} \cdot {\bf S}_{j,n+2} \rangle 
\right),
\nonumber \\
\label{eq:CDoperator}
\end{eqnarray}
where $\langle \cdots \rangle$ denotes the expectation value 
in the ground state, i.e., the lowest-energy state in the subspace 
of zero magnetization 
$S^z_{\rm tot} = \sum_{j,n} S^z_{j,n} = 0$.
In the CD phase, the CD order induced at open boundaries of the ladder 
penetrates into the bulk and exhibits a long-range order.
In the other phases with a spin gap, the CD order is expected 
to decay exponentially when we move from the boundary into the bulk, 
while we expect that the CD order decays algebraically at a critical point.
We may therefore be able to identify the CD phase and transition points 
by monitoring the system-size dependence of the CD operator $D_{\rm CD}(n; L)$ 
at the center of open ladder, $n=L/2$.
In the calculation, we set $L$ to be a multiple of four so that 
$D_{\rm CD}(L/2; L)$ is positive.

\begin{figure}
\begin{center}
\includegraphics[width=0.4\textwidth]{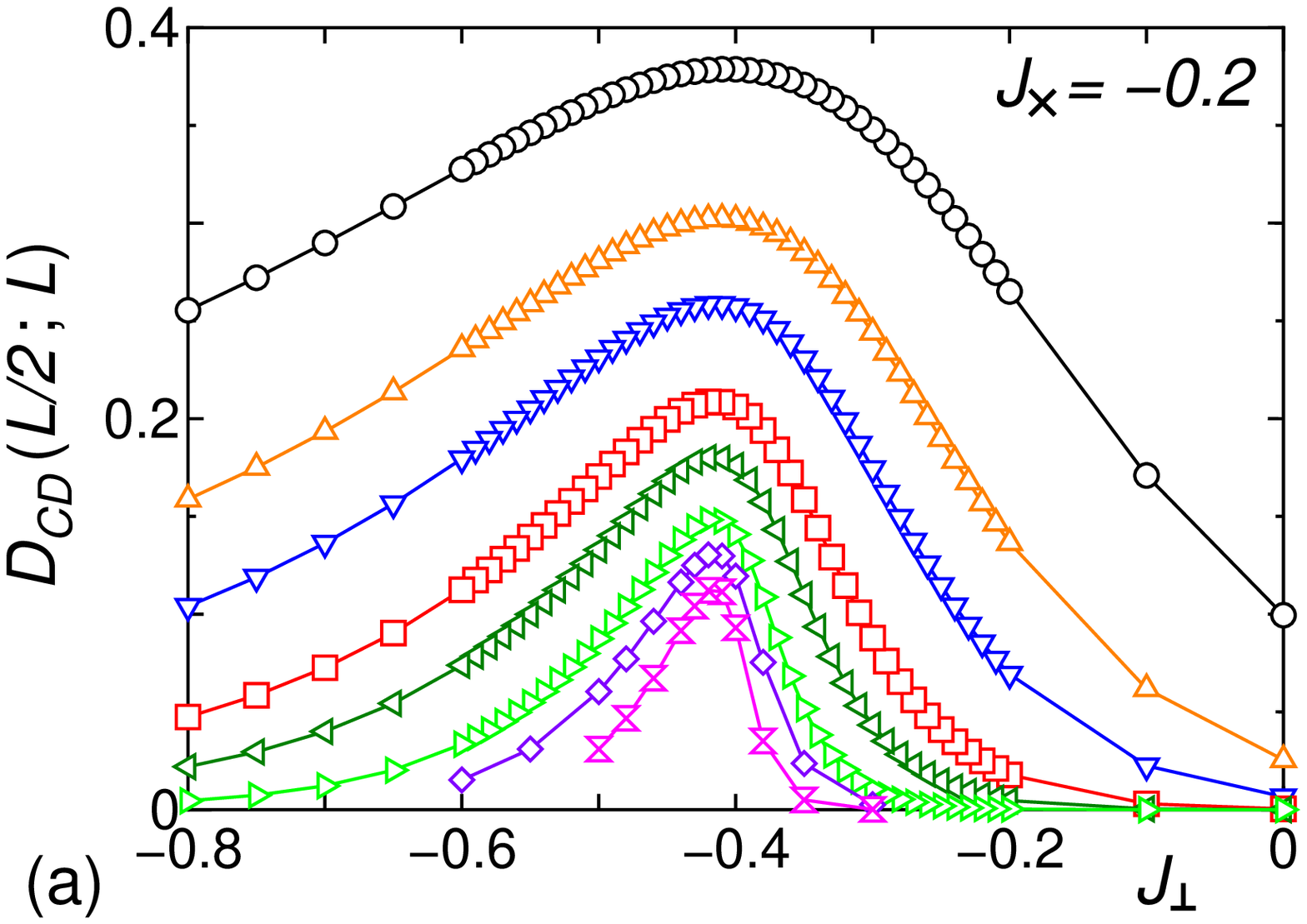}
\includegraphics[width=0.4\textwidth]{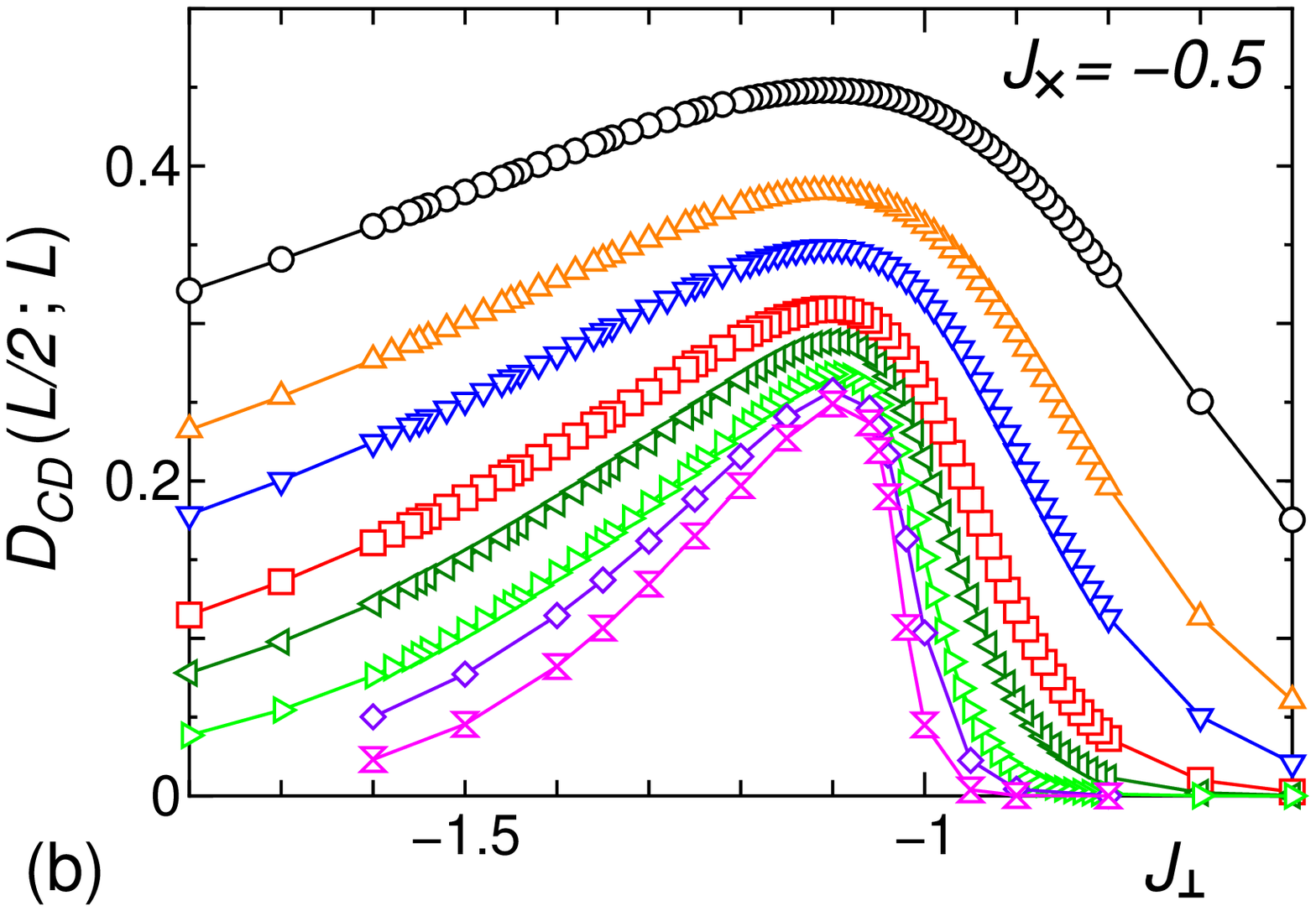}
\includegraphics[width=0.4\textwidth]{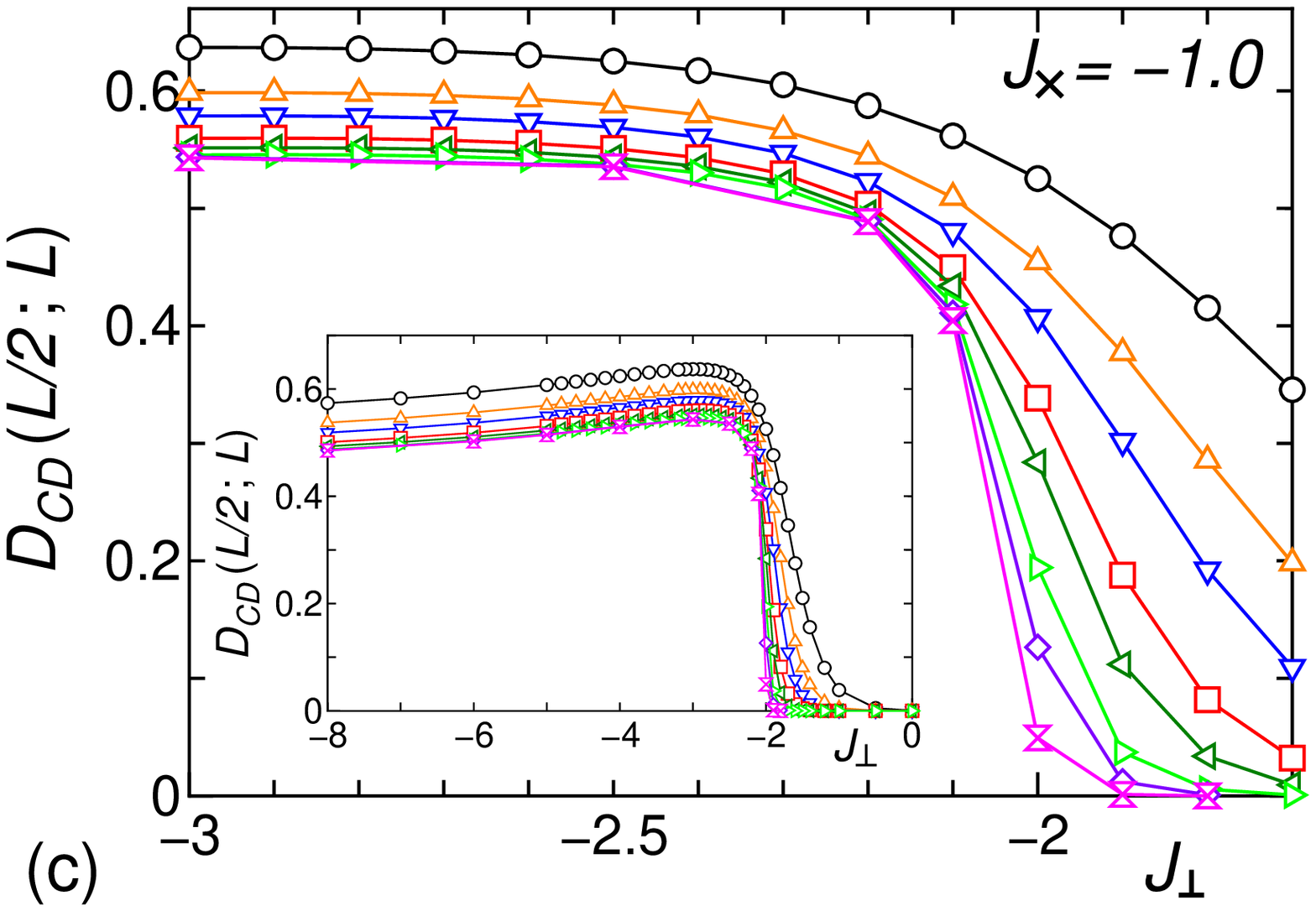}
\end{center}
\caption{
(Color online) 
CD operator at the center of the open ladder, $D_{\rm CD}(L/2; L)$ 
as a function of $J_\perp$ 
for $J = 1$ and 
(a) $J_\times = -0.2$, (b) $J_\times = -0.5$, and
(c) $J_\times = -1.0$.
The symbols represent the data 
for $L=16, 24, 32, 48, 64, 96, 128$ and $192$ from top to bottom.
Inset in (c) shows the data for $J_\times = -1.0$ 
and broader regime of $J_\perp$, $-8 \le J_\perp \le 0$.
}
\label{fig:Dcd-Jp}
\end{figure}

\begin{figure}
\begin{center}
\includegraphics[width=0.36\textwidth]{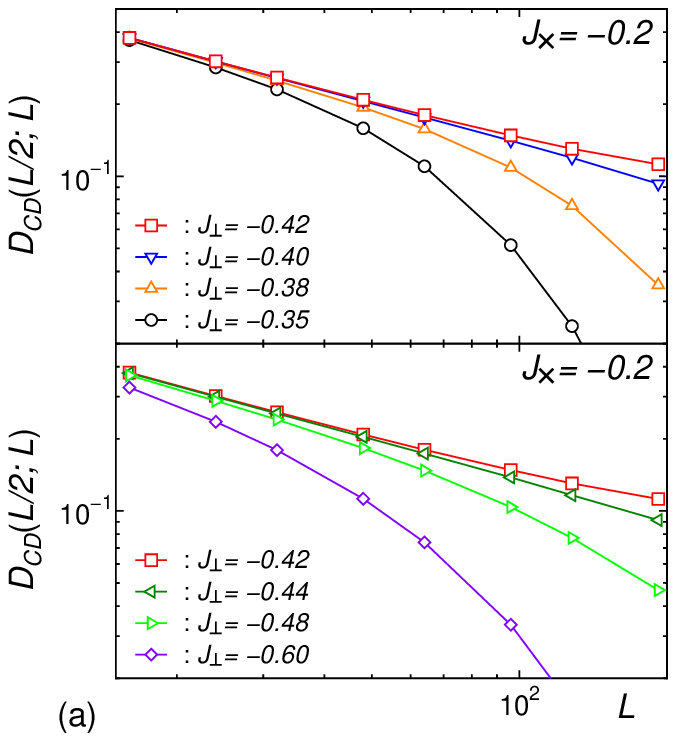}
\includegraphics[width=0.36\textwidth]{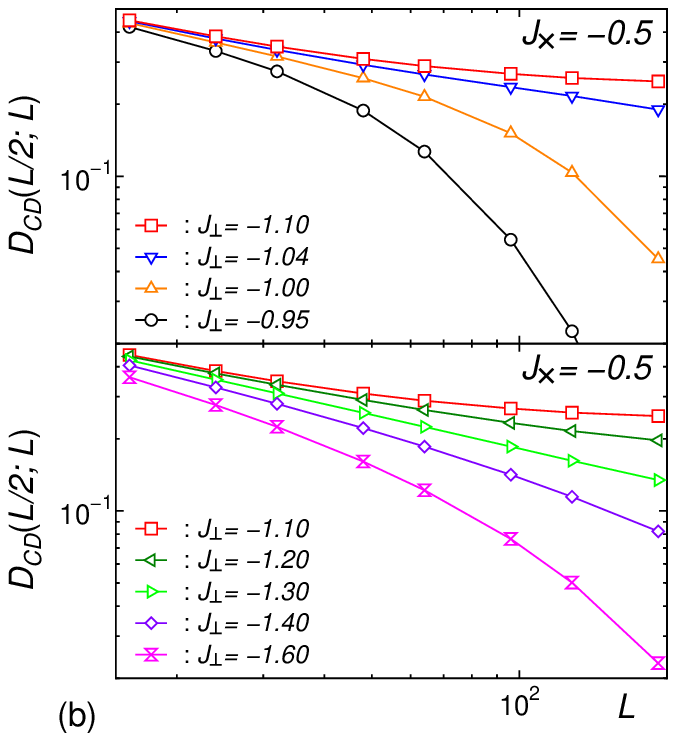}
\includegraphics[width=0.36\textwidth]{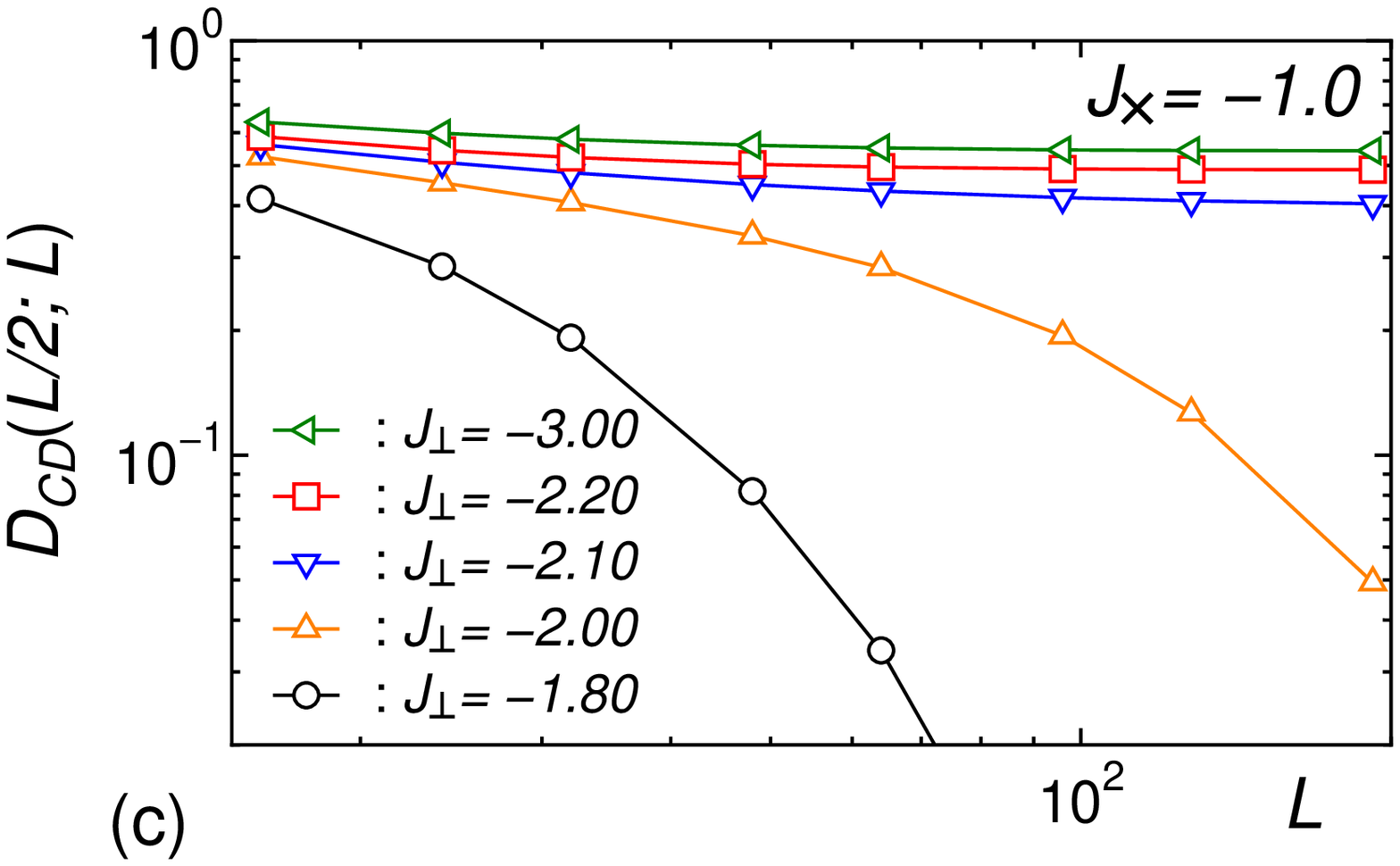}
\end{center}
\caption{
(Color online) 
System-size dependence of the CD operator at the center of the open ladder, 
$D_{\rm CD}(L/2; L)$, in a log-log scale 
for $J = 1$ and (a) $J_\times = -0.2$, (b) $J_\times = -0.5$, 
and (c) $J_\times = -1.0$.
}
\label{fig:Dcd-L}
\end{figure}

\begin{figure}
\begin{center}
\includegraphics[width=0.45\textwidth]{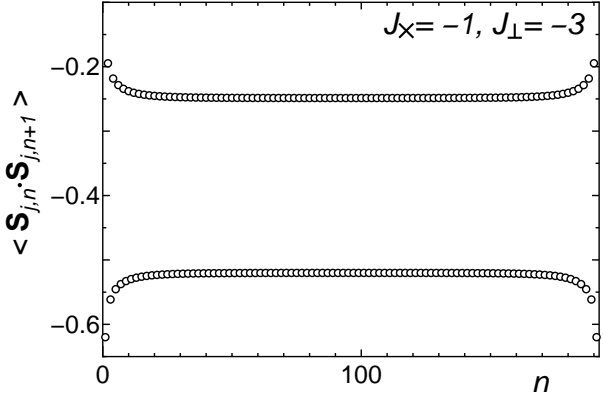}
\end{center}
\caption{
Spin correlations between nearest-neighbor spins
(bond energy) along the legs of the ladder, 
$\langle {\bf S}_{j,n} \cdot {\bf S}_{j,n+1} \rangle$, 
 for $(J, J_\times, J_\perp) = (1, -1, -3)$ and $L=192$.
The correlations in the legs $j=1$ and $j=2$ are identical.
}
\label{fig:NNcor}
\end{figure}

\begin{figure}
\begin{center}
\includegraphics[width=0.36\textwidth]{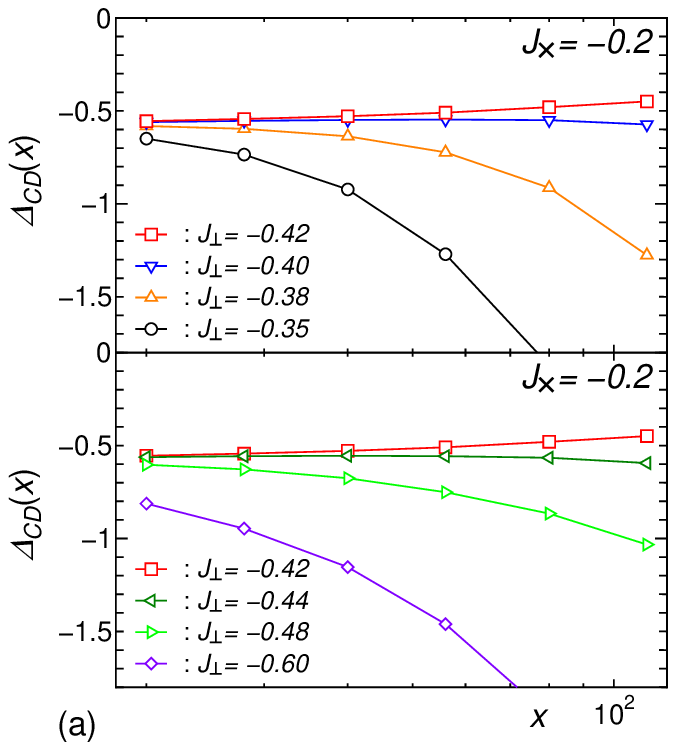}
\includegraphics[width=0.36\textwidth]{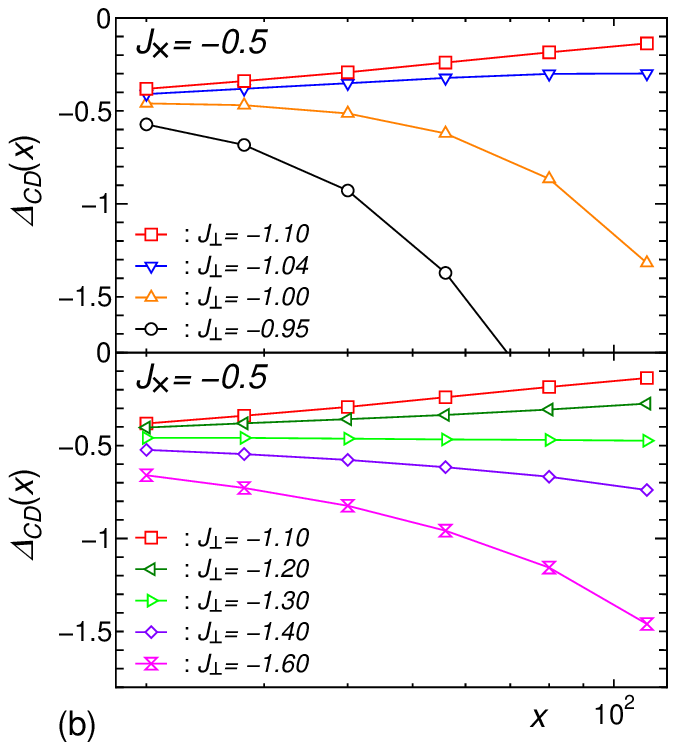}
\includegraphics[width=0.36\textwidth]{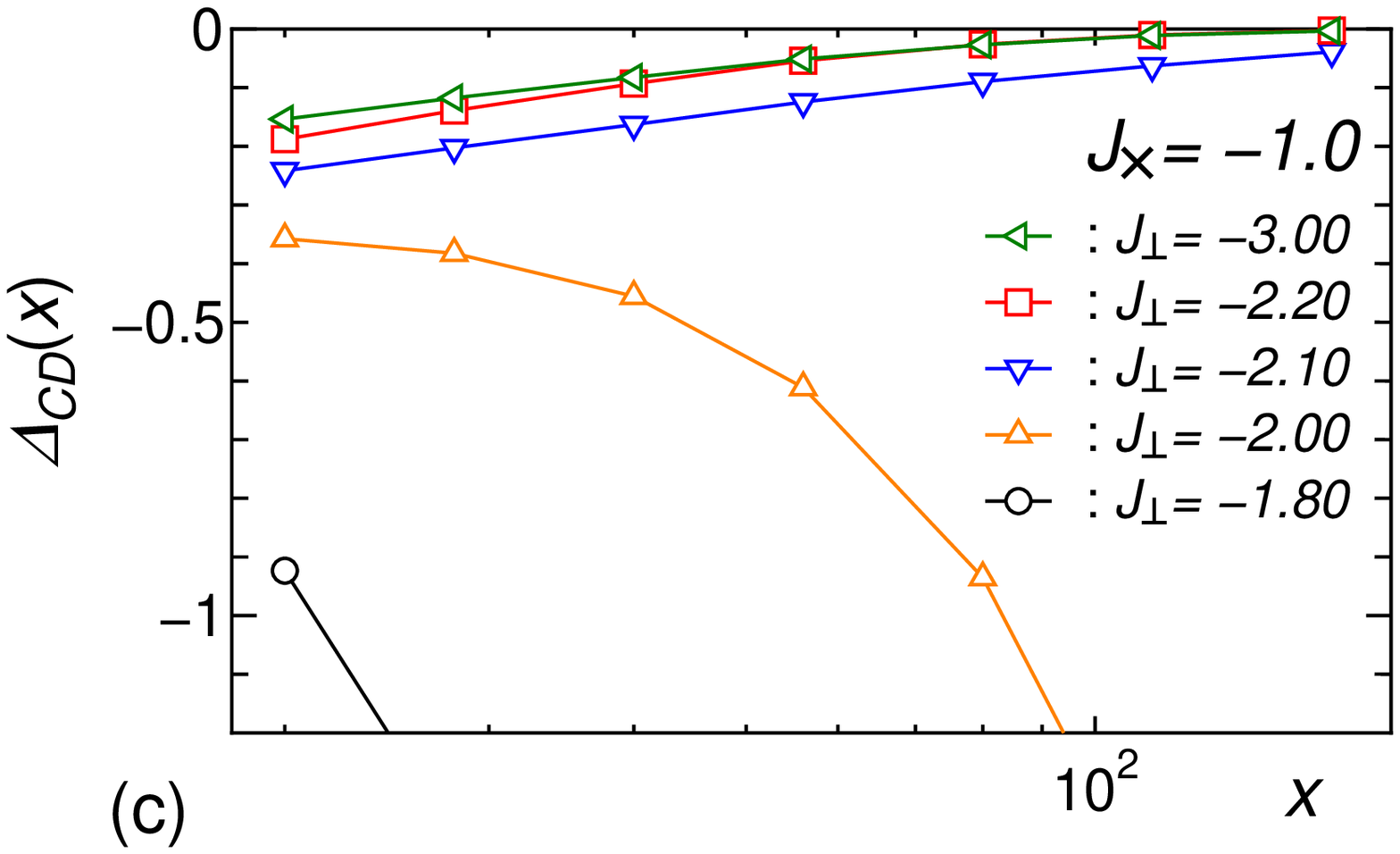}
\end{center}
\caption{
(Color online) 
Slope of the $\log D_{\rm CD}(L/2; L)$-$\log L$ plot, 
$\Delta_{\rm CD}(x)$, for $J = 1$ and 
(a) $J_\times = -0.2$, (b) $J_\times = -0.5$, 
and (c) $J_\times = -1.0$.
}
\label{fig:Deltacd-L}
\end{figure}

Figure\ \ref{fig:Dcd-Jp} shows the dependence of the CD operator 
$D_{\rm CD}(L/2; L)$ on the rung coupling $J_\perp$ 
for $J = 1$ and several fixed $J_\times$.
We find that $D_{\rm CD}(L/2; L)$ has a broad peak, indicating that 
the CD order is strong in a rather wide regime of $J_\perp$.
We note that for $J_\times = -1$ and $J_\perp \lesssim -2$, 
the $L$ convergence of $D_{\rm CD}(L/2; L)$ seems almost achieved, 
suggesting the appearance of the CD long-range order.

In order to determine whether or not the CD order survives 
in the thermodynamic limit, we investigate the system-size dependence 
of the CD operator $D_{\rm CD}(L/2; L)$.
Figure\ \ref{fig:Dcd-L} shows 
the $L$ dependence of the CD operator $D_{\rm CD}(L/2; L)$ 
for some typical sets of coupling parameters.
It is clear that $D_{\rm CD}(L/2; L)$ for $J_\times=-1$ 
and large negative $J_\perp$ converges to a finite value at $L \to \infty$.

We also show in Fig.\ \ref{fig:NNcor} the spatial profile of 
the nearest-neighbor spin correlations 
$\langle {\bf S}_{j,n} \cdot {\bf S}_{j,n+1} \rangle  = C_{\rm av} + (-1)^n \langle \epsilon_n \rangle$ 
for $(J, J_\times, J_\perp) = (1, -1, -3)$, 
which clearly demonstrates the presence of well-developed columnar dimer order.
The average energy density, calculated in the middle of the ladder, is found to be
$C_{\rm av} = (\langle {\bf S}_{j,L/2-1} \cdot {\bf S}_{j,L/2} \rangle
+ \langle {\bf S}_{j,L/2} \cdot {\bf S}_{j,L/2+1} \rangle)/2 = -0.384$.
We see strong modulation of the bond energy $\langle {\bf S}_{j,n} \cdot {\bf S}_{j,n+1} \rangle$
 between even and odd bonds. 
The amplitude of the modulation saturates in the middle of the ladder where 
the bulk dimerization value is achieved, $\langle \epsilon_n \rangle \to 0.136$.
We find that bond modulations in the two chains are in-phase,
implying {\em columnar} ordering of stronger and weaker bonds.
This finding represents direct proof of the 
CD phase in the frustrated ladder model (\ref{eq:Ham}) with ferromagnetic inter-chain exchanges.

For smaller $|J_\times|$, on the other hand, 
the appearance of the CD long-range order is not so clear; 
$D_{\rm CD}(L/2; L)$ still decreases with $L$ 
even at the largest $L$ calculated [see Fig.\ \ref{fig:Dcd-L} (a) and (b)].
However, we find that in some parameter regime $D_{\rm CD}(L/2; L)$  
{\em bends upward} in a log-log plot. 
This means that the decay of $D_{\rm CD}(L/2; L)$ becomes slower 
as $L$ gets larger, which suggests the emergence of the CD long-range order 
in the thermodynamic limit.

To elucidate the bending-up behavior, we also investigate 
the system-size dependence of the slope of the log-log plot,
\begin{eqnarray}
&& \Delta_{\rm CD}(x_i)  \nonumber \\
&&=\frac{\log[D_{\rm CD}(L_{i+1}/2; L_{i+1})]-\log[D_{\rm CD}(L_i/2; L_i)]}
{\log(L_{i+1})-\log(L_i)}, \nonumber \\
\label{eq:Deltacd-x}
\end{eqnarray}
where $x_i = (L_i+L_{i+1})/2$ and $L_i=16,24,32,48,64,96,128,192$ 
for $i=1,2,...,8$.
If $D_{\rm CD}(L/2; L)$ decays exponentially with increasing $L$, 
the slope $\Delta_{\rm CD}(x)$ decreases as $x$ increases.
If $D_{\rm CD}(L/2; L)$ exhibits a long-range order, 
$\Delta_{\rm CD}(x)$ increases with $x$ and converges to zero at $x \to \infty$.
Furthermore, if $D_{\rm CD}(L/2; L)$ decays algebraically, 
$\Delta_{\rm CD}(x)$ converges to a finite negative value at $x \to \infty$.
Figure\ \ref{fig:Deltacd-L} shows the data of $\Delta_{\rm CD}(x)$ 
as a function of $x$.\cite{convergence-of-slope}
The results clearly suggest that there are parameter regions 
where $\Delta_{\rm CD}(x)$ increases with $x$.
We take this behavior as an evidence of the CD phase.

Based on the above results we conclude that the CD phase emerges 
in a finite region in the $J_\perp$ - $J_\times$ plane.
The phase boundaries estimated from the results of the slope 
$\Delta_{\rm CD}(x)$ above are plotted in the phase diagram, see
Fig.\ \ref{fig:phasediagram} in Sec.\ \ref{subsection:phasediagram}.
We note that, as shown in the Appendix, 
the bending-up behavior of the dimer operator in the log-log plots 
is also observed 
in the frustrated Heisenberg chain (which can also be viewed as the 
zigzag ladder), which is well known 
to exhibit the dimer phase for sufficiently large next-nearest-neighbor 
exchange $J_2$.\cite{MajumdarG1969A,MajumdarG1969B,Haldane1982,JullienH1983,OkamotoN1992}
This observation provides us with an important check of the approach to the frustrated ladder (\ref{eq:Ham})
and supports  our interpretation of the data in Figs.\ \ref{fig:Dcd-L} and \ref{fig:Deltacd-L}.

\begin{figure}
\begin{center}
\includegraphics[width=0.36\textwidth]{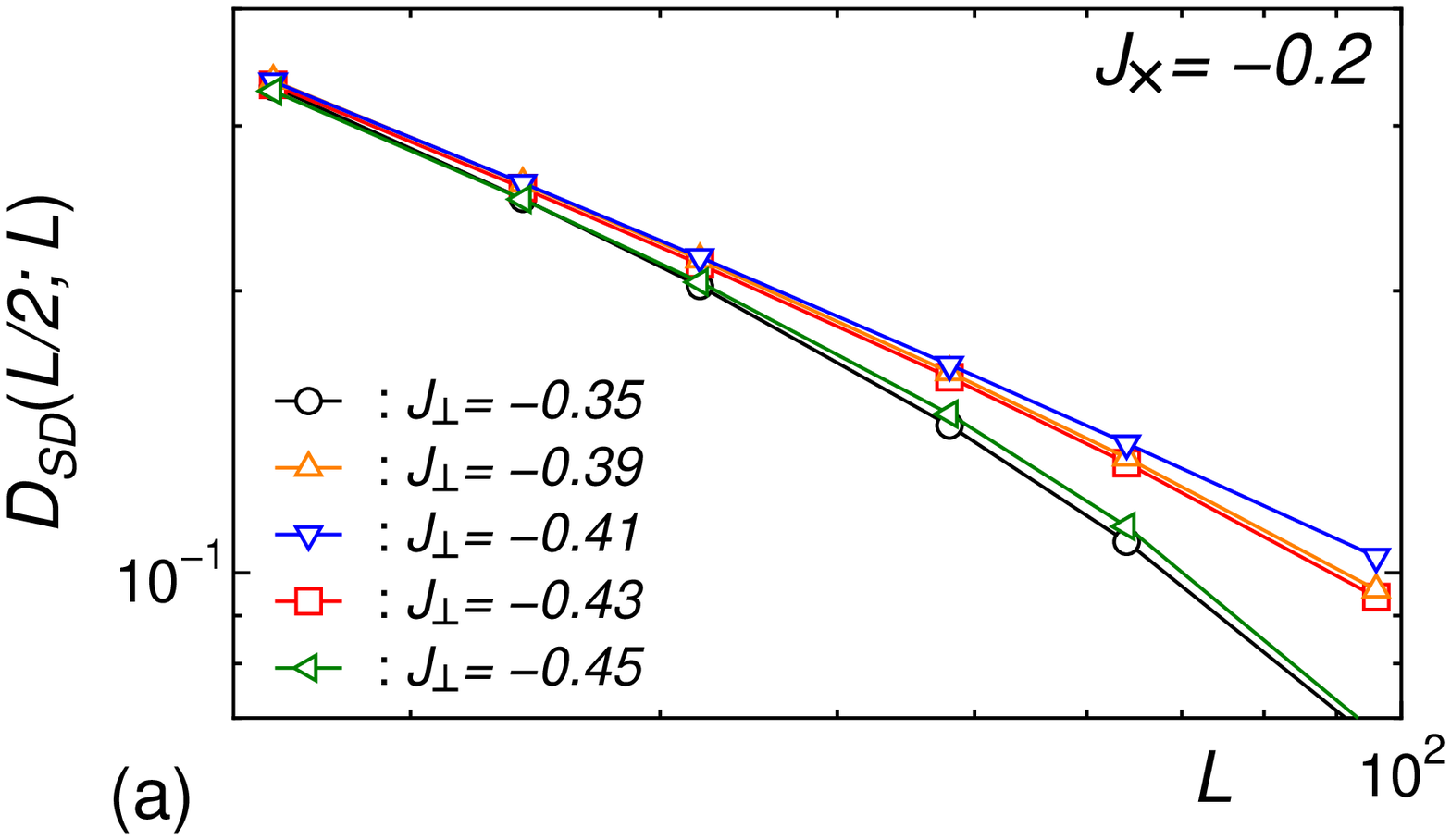}
\includegraphics[width=0.36\textwidth]{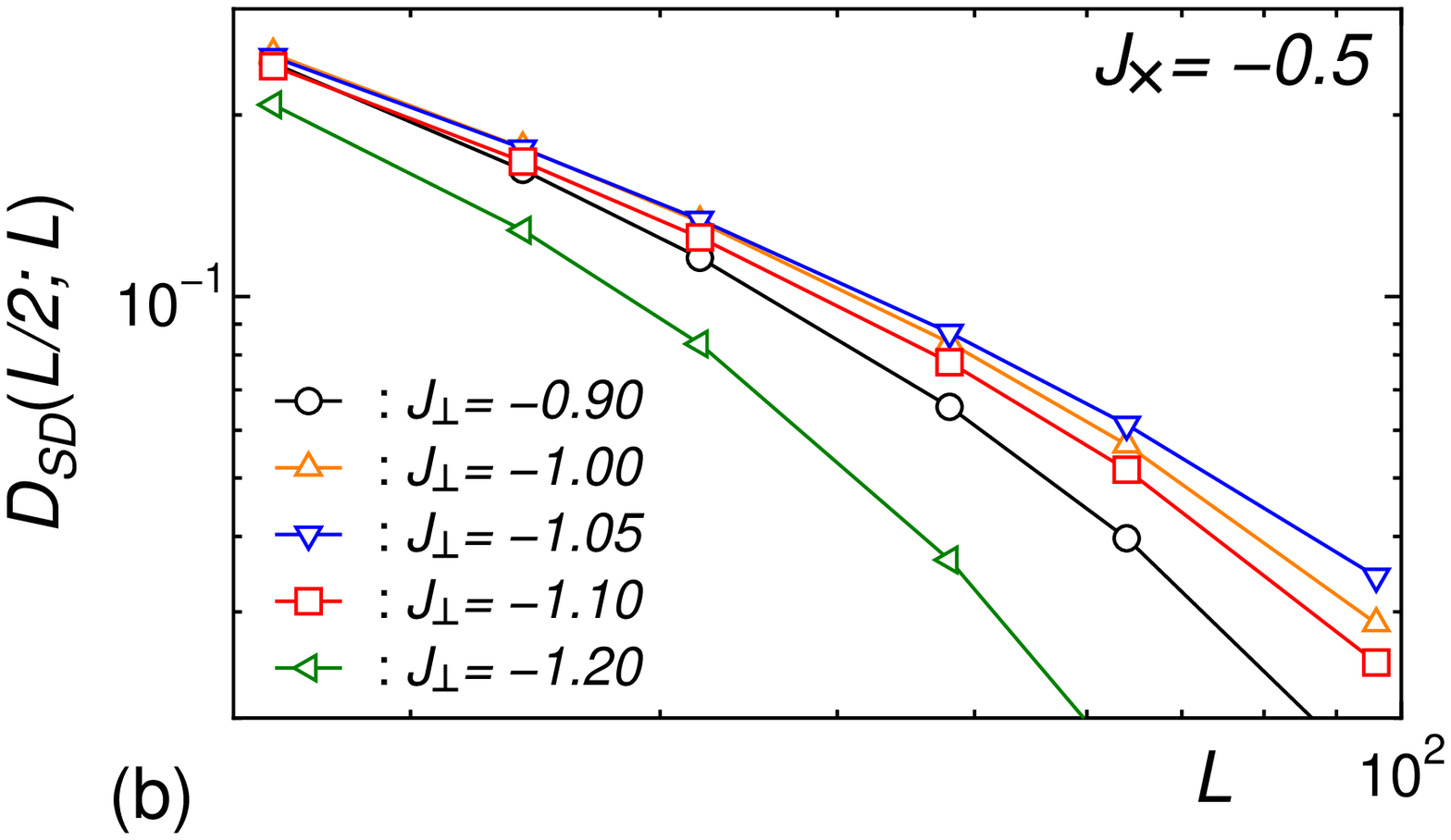}
\includegraphics[width=0.36\textwidth]{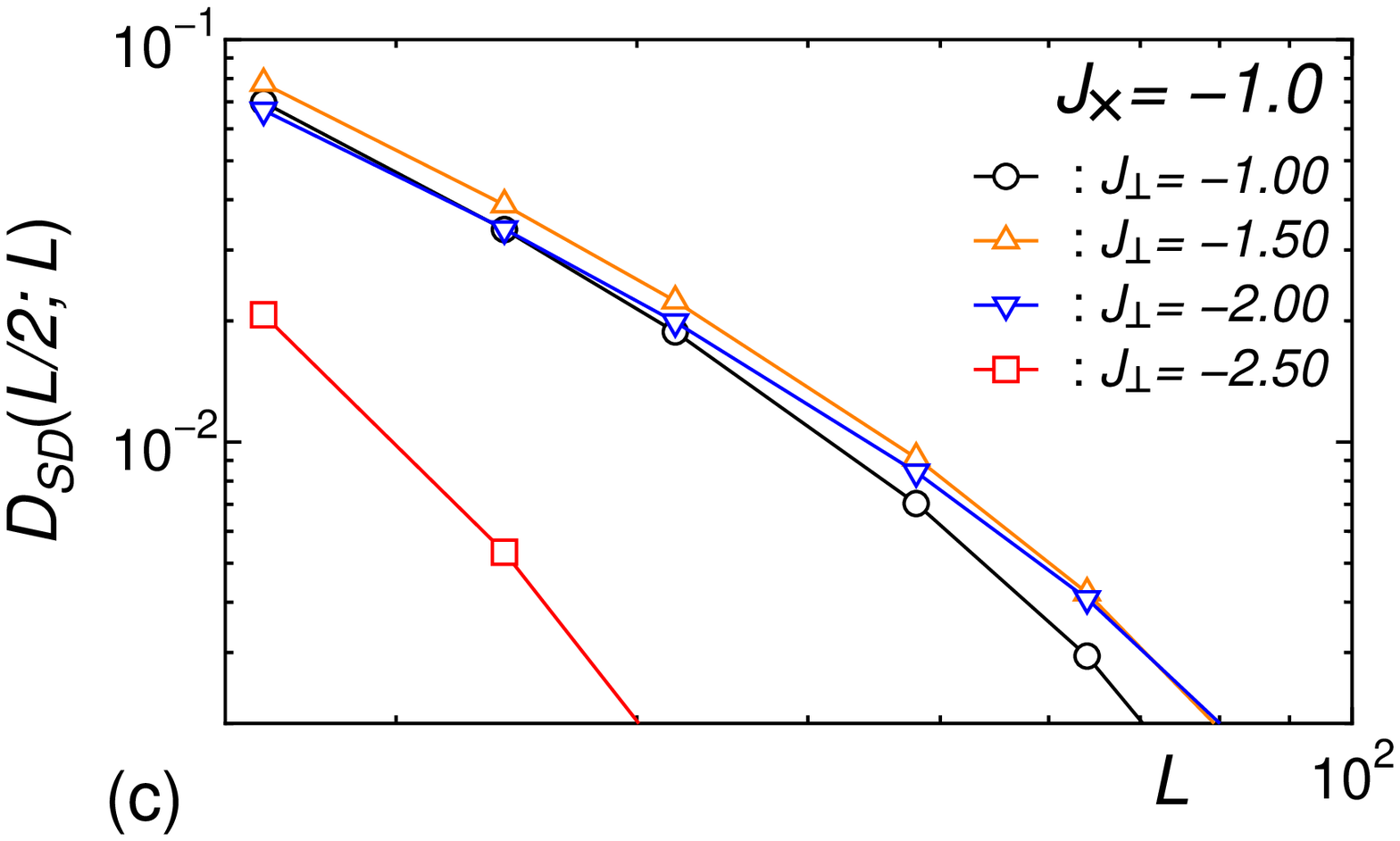}
\end{center}
\caption{
(Color online) 
System-size dependence of the SD operator at the center of the open ladder, 
$D_{\rm SD}(L/2; L)$, in a log-log scale 
for $J = 1$ and (a) $J_\times = -0.2$, (b) $J_\times = -0.5$, 
and (c) $J_\times = -1.0$.
}
\label{fig:Dsd-L}
\end{figure}

Although not expected from the RG analysis, 
we have also examined possibility of the SD order in the model.
For this purpose, we have calculated the local SD operator,
\begin{eqnarray}
&&D_{\rm SD}(n; L) \nonumber \\
&& = \sum_{j=1,2} (-1)^j \left(
\langle {\bf S}_{j,n} \cdot {\bf S}_{j,n+1} \rangle 
- \langle {\bf S}_{j,n+1} \cdot {\bf S}_{j,n+2} \rangle \right), \nonumber \\
\end{eqnarray}
in the frustrated ladder (\ref{eq:Ham}) with up to $L=96$ rungs.
In the calculation of the SD operator, we have employed 
an open boundary condition with an extra spin at each edge, 
which selects one of the SD patterns and lifts the two-fold degeneracy 
in the possible SD ground states.
The results are presented in Fig.\ \ref{fig:Dsd-L}.
We find that $D_{\rm SD}(L/2; L)$ bends downward in a log-log plot, 
indicating the exponential decay.
[$D_{\rm SD}(L/2; L)$ for $J_\times=-0.2$ and $J_\perp=-0.41$, 
at which point we have found that the decay of the SD order is the slowest, 
exhibits a nearly-linear behavior, but it actually bends down slightly.]
We have performed the same calculation for a wide parameter regime 
and found that $D_{\rm SD}(L/2; L)$ decays exponentially in each gapped phase 
or, at most, decays algebraically at a transition point.
We thus conclude that the SD phase is absent 
in the model (\ref{eq:Ham}) with ferromagnetic $J_\perp$ and $J_\times$.

\subsection{$z$ operator}
\label{subsection:z-op}

Here, we discuss another numerical approach to the problem, based on
so-called ``$z$ operators",\cite{NakamuraT2002,NakamuraT2002B}
which are used to distinguish
different valence-bond-solid (VBS) states in one-dimensional spin systems.
For the frustrated ladder model (\ref{eq:Ham}), two $z$ operators, 
$z_{\rm rung}$ and $z_{\rm diag}$, are defined as follows,
\begin{eqnarray}
z_{\rm rung}(L) &=& 
\langle \exp\left[ i \frac{2\pi}{L} \sum_{n=1}^L 
n (S^z_{1,n} + S^z_{2,n}) \right] \rangle,
\nonumber \\
z_{\rm diag}(L) &=& 
\langle \exp\left[ i \frac{2\pi}{L} \sum_{n=1}^L 
n (S^z_{1,n+1} + S^z_{2,n}) \right] \rangle.
\label{eq:z-operator}
\end{eqnarray}
It has been shown \cite{NakamuraT2002,NakamuraT2002B} that the $z$ operators 
in the spin-1/2 two-leg ladder with $L$ rungs 
under the periodic boundary condition exhibits the following asymptotic behavior with $L$, 
\begin{eqnarray}
z_{\rm rung/diag}(L) \sim (-1)^{N_{\rm VBS}} \left[ 1 - \mathcal{O}(1/L) \right],
\label{eq:z-asymp}
\end{eqnarray}
where $N_{\rm VBS}$ is an integer depending on 
the VBS pattern of the state under consideration:
it represents the number of singlet bonds `cut' by a line parallel to the rung/diagonal link.
The $z$ operator then measures topological parity of the dimer covering pattern
describing particular gapped state. 
In our case, $z_{\rm rung}$ converges to $1$ for the RS and CD states 
(even number of singlets crossed) in the thermodynamic limit, 
while $z_{\rm rung} \to -1$ for the Haldane state (the number of crossed singlets is always odd).
Conversely, $z_{\rm diag} \to -1$ for the RS and CD states, while $z_{\rm diag} \to 1$ in the Haldane state. 
A remarkable feature of the $z$ operators is that 
they change their sign at the transition between phases 
having different parity of $N_{\rm VBS}$.
This property makes the $z$ operators more powerful in 
determining the critical point of such a phase transition 
than the string order parameter, which just vanishes at the transition.\cite{kim00}
Indeed, the $z$ operators have turned out to be successful in determining 
the direct RS-Haldane transition point occurring for large antiferromagnetic 
$J_{\perp,\times}$.\cite{NakamuraT2002B} 
For the present case of ferromagnetic $J_{\perp,\times}$, 
we can use $z_{\rm rung}$ and $z_{\rm diag}$
to locate the transition point between the CD and Haldane phases.

\begin{figure}
\begin{center}
\includegraphics[width=0.49\textwidth]{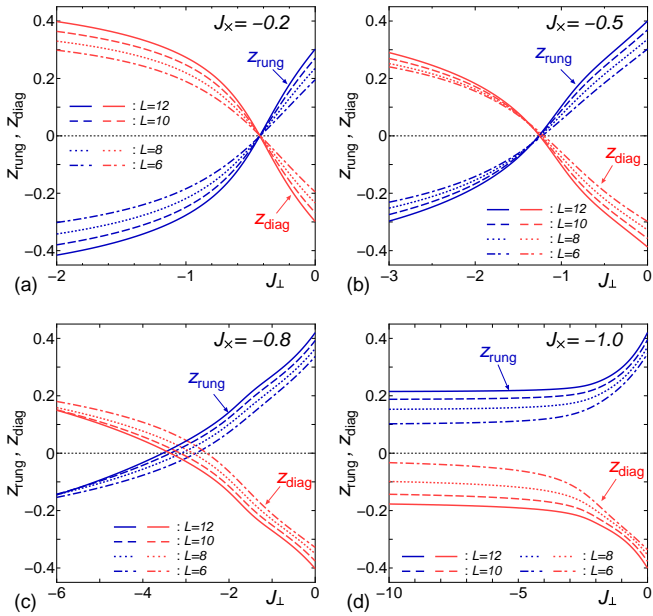}
\end{center}
\caption{
(Color online) 
$J_\perp$ dependence of the $z$ operators 
for $J = 1$ and 
(a) $J_\times = -0.2$, (b) $J_\times = -0.5$, 
(c) $J_\times = -0.8$, (d) $J_\times = -1.0$.
Dark (blue) and light (red) curves represent 
$z_{\rm rung}$ and $z_{\rm diag}$, respectively.
}
\label{fig:z-opr}
\end{figure}

Using the exact-diagonalization method, we have calculated  
the $z$ operators, $z_{\rm rung}$ and $z_{\rm diag}$, 
in the ladder (\ref{eq:Ham}) with up to $L = 12$ rungs 
under the periodic boundary condition.
Figure\ \ref{fig:z-opr} presents the results 
for typical parameter lines with $J=1$ and fixed $J_\times$.
For $J_\times < 1$, we have observed the sign change in $z_{\rm rung}$ 
($z_{\rm diag}$) from positive (negative) to negative (positive) values 
as $J_\perp$ decreases.
The crossing point of $z_{\rm rung}$ and $z_{\rm diag}$ thereby gives 
an estimate of the transition point between the CD and Haldane phases.
While the $L$ dependence is negligibly small for small $|J_\times|$, 
the crossing point for large $|J_\times|$ moves sizably with $L$, 
suggesting that the finite-size effects still remain.
However, we emphasize that the crossing point shifts towards 
smaller $J_\perp$ with increasing $L$, which means that 
the range of CD phase {\sl broadens} as $L$ increases, 
and approaches smoothly to the CD-Haldane transition point 
obtained from the DMRG analysis.
(See also the phase diagram, Fig.\ \ref{fig:phasediagram} 
in Sec.\ \ref{subsection:phasediagram}.)
Thus, we can safely state that the analysis of $z$ operators 
also supports the appearance of the CD phase.
For $J_\times = 1$, $z_{\rm rung}$ ($z_{\rm diag}$) is positive (negative) 
for the entire regime of $J_\perp$ calculated.
The result is consistent with the prediction 
of the perturbative analysis in Sec.\ \ref{subsection:perturb} 
as well as the DMRG results in Sec.\ \ref{subsection:dmrg}, 
which show that the CD phase extends to the limit $J_\perp \to -\infty$.

\subsection{Phase diagram}
\label{subsection:phasediagram}

\begin{figure}
\begin{center}
\includegraphics[width=0.45\textwidth]{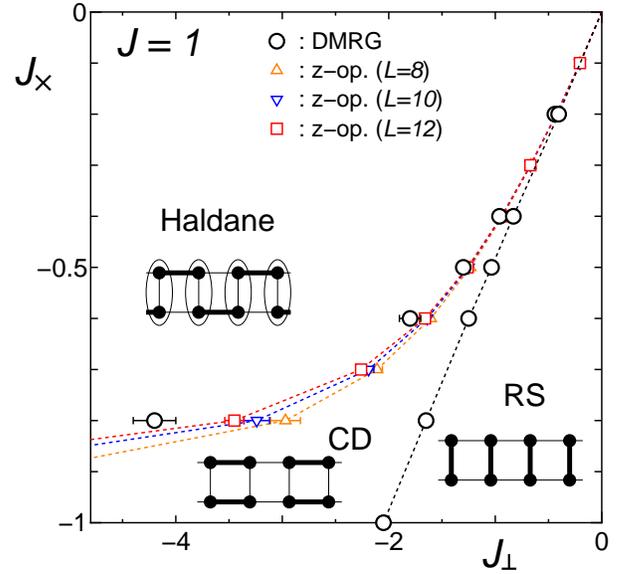}
\end{center}
\caption{
(Color online) 
Ground-state phase diagram for $J=1$ and 
the ferromagnetic interchain coupling, $J_\perp < 0$ and $J_\times < 0$.
Open circles represent the transition points obtained from 
the DMRG calculation
while the other symbols show the CD-Haldane transition points 
from the analysis of $z$ operators.
Dotted lines are guide to eye.
In the schematic pictures for each phase, the bold lines represent 
singlet pairs and the ellipses stand for the symmetrization of two spins.
}
\label{fig:phasediagram}
\end{figure}

Combining the above results, we determine the ground-state phase diagram 
in the parameter plane for ferromagnetic $J_\perp$ and $J_\times$.
Figure\ \ref{fig:phasediagram} shows the resultant phase diagram, 
which includes the Haldane, RS, and CD phases.
We clearly see that the CD phase appears in a wide parameter region,
which is seen to expand as $|J_\perp|$ and $|J_\times|$ become bigger.
The transition line between the RS and CD phases 
seems to nearly coincide with the line of $J_\perp = 2 J_\times$.
The boundary between the Haldane and CD phases 
starts from $J_\perp = J_\times = 0$ and runs towards smaller $J_\perp$ 
as $J_\times$ decreases, approaching smoothly 
the limit of the strong rung-exchange, 
$J_\times = -J$ at $J_\perp \to -\infty$.
It is worth noting that the DMRG result on the RS-CD transition line 
agrees even quantitatively with the result of RG analysis 
in Table \ref{table1}, 
and the behavior of the CD-Haldane transition line 
is also consistent with the analytical RG result.
This observation strongly supports the correctness of 
the RG analysis in Sec.\ \ref{section:rg}.

\section{Antiferromagnetic Inter-chain couplings}
\label{section:AF}

The numerical results in Sec.\ \ref{section:FM} have revealed 
that the frustrated ladder (\ref{eq:Ham}) 
with ferromagnetic $J_\perp$ and $J_\times$ exhibits 
the CD phase in a wide parameter regime, in agreement
with the prediction of RG analysis in Sec.\ \ref{section:rg}.
Since the validity of the RG analysis relies only on the small amplitudes
of the inter-chain couplings $J_\perp$ and $J_\times$ 
and is not affected by their signs, 
we naturally expect that the RG analysis is correct also 
for the antiferromagnetic couplings.
To examine the expectation, we re-visit the frustrated ladder (\ref{eq:Ham}) 
with antiferromagnetic $J_\perp$ and $J_\times$.
For this case, it has been shown rather clearly that 
for large $J_\perp$ and $J_\times$ 
the direct first-order transition takes place 
between the RS and Haldane phases,\cite{kim08}
while the situation is still controversial 
for small $J_\perp$ and $J_\times$.\cite{hung06,kim08,liu08}
To clarify the situation we have performed the DMRG calculation for a parameter 
line $J=1$ and $J_\times=0.2$ 
and investigated behaviors of the CD and SD operators.

\begin{figure}
\begin{center}
\includegraphics[width=0.36\textwidth]{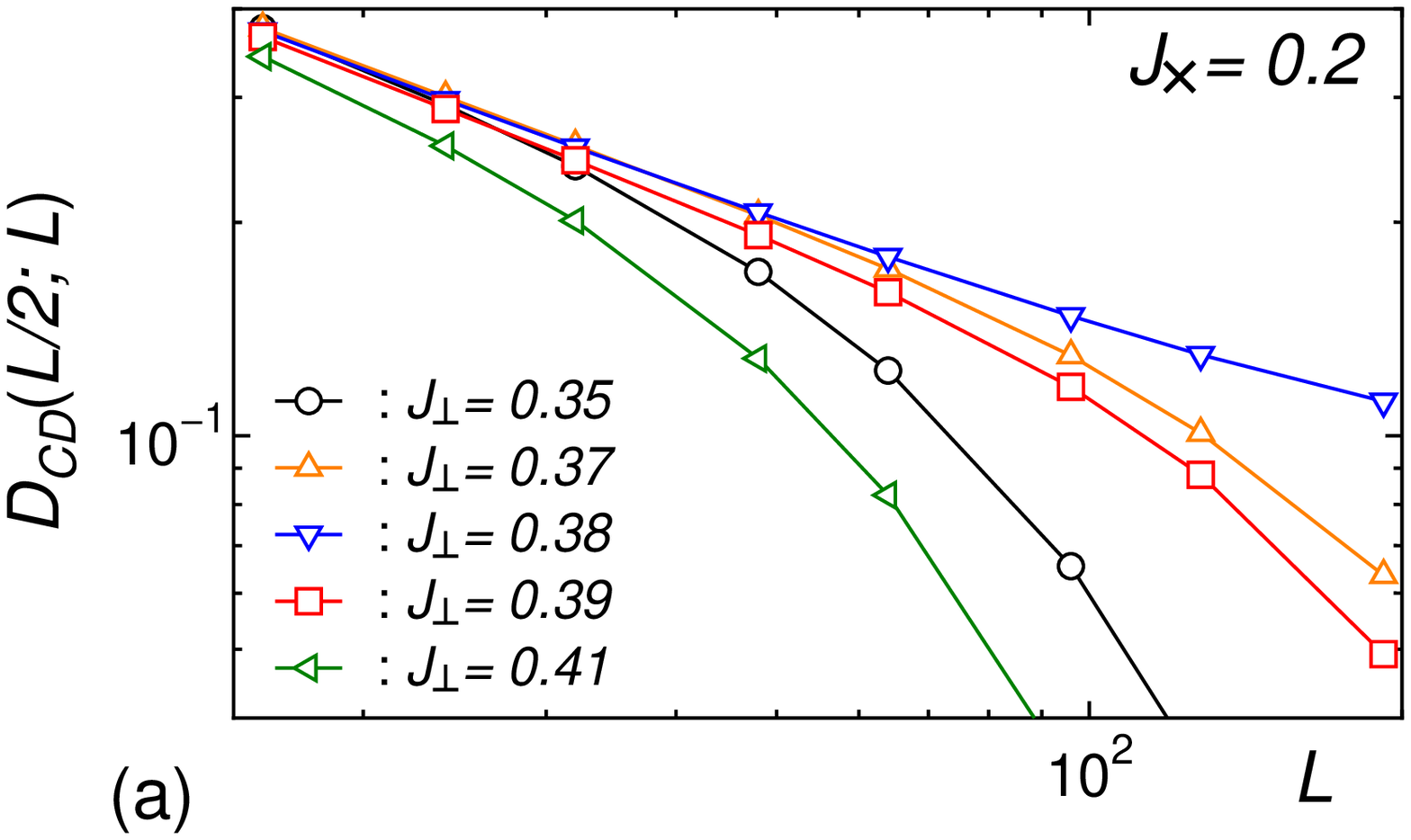}
\includegraphics[width=0.36\textwidth]{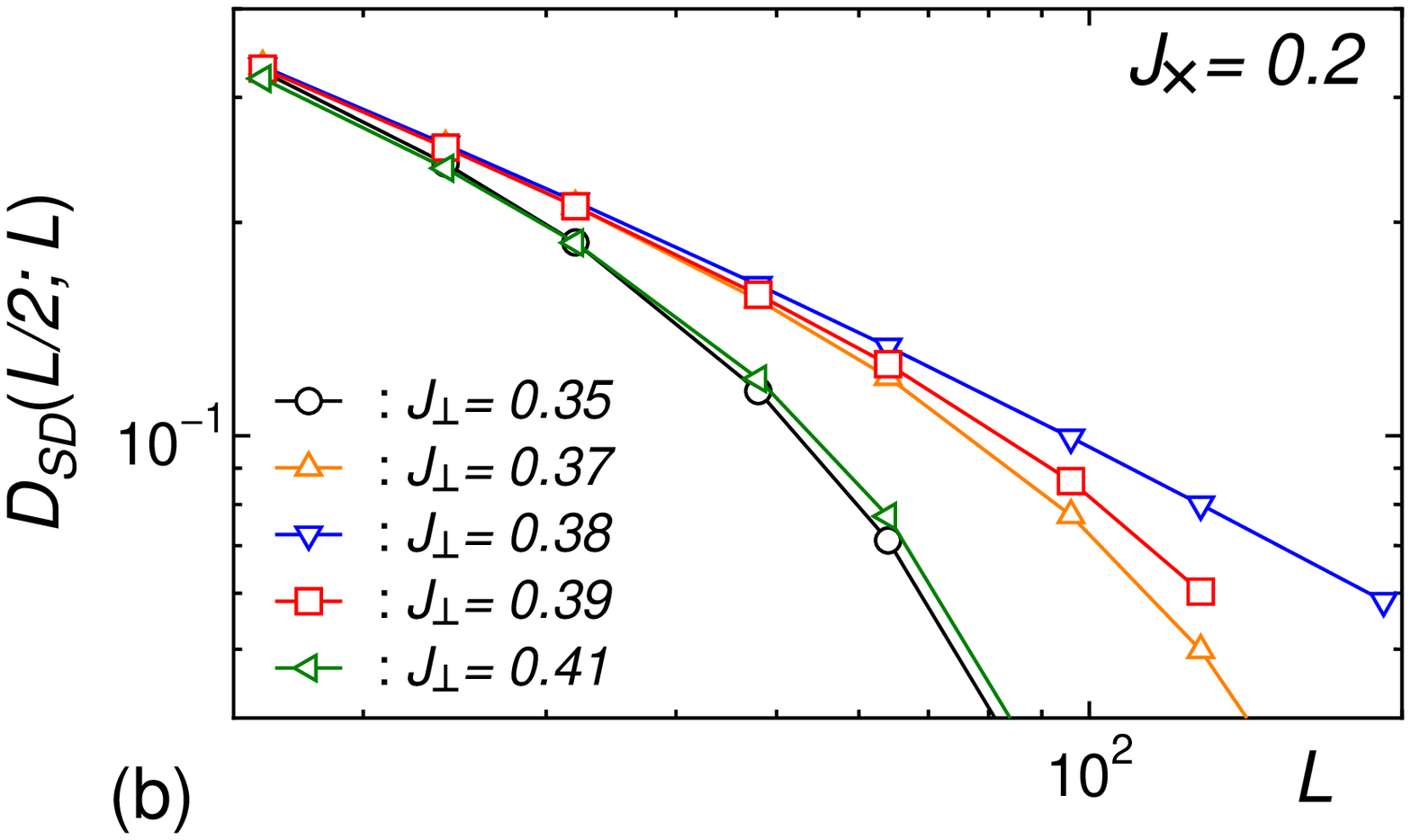}
\end{center}
\caption{
(Color online) 
System-size dependence of the dimer operators 
at the center of the open ladder in a log-log scale for $J = 1$ and 
$J_\times = 0.2$;
(a) the CD operator $D_{\rm CD}(L/2; L)$ 
and (b) the SD operator $D_{\rm SD}(L/2; L)$.
}
\label{fig:D-L-AF}
\end{figure}

\begin{figure}
\begin{center}
\includegraphics[width=0.36\textwidth]{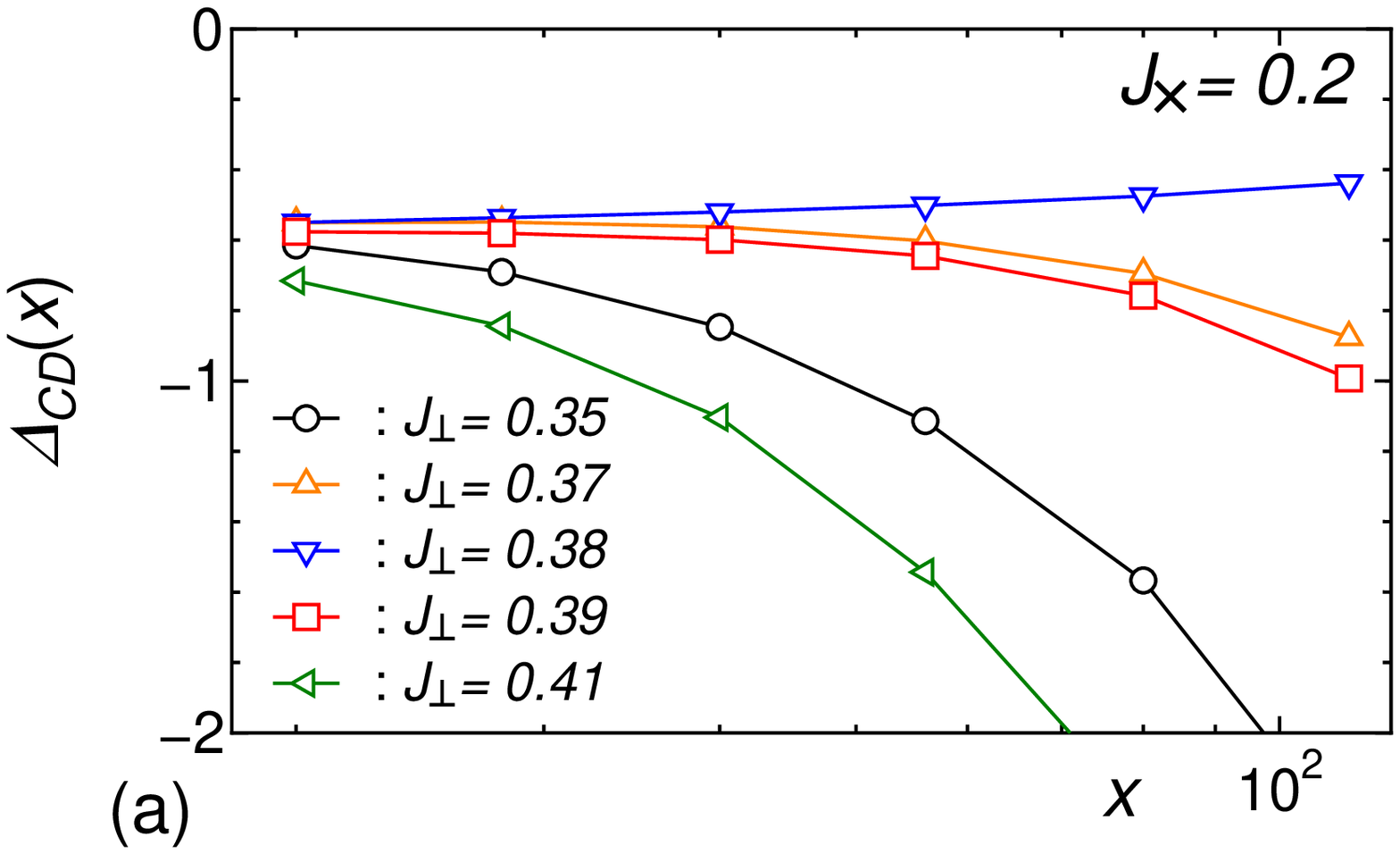}
\includegraphics[width=0.36\textwidth]{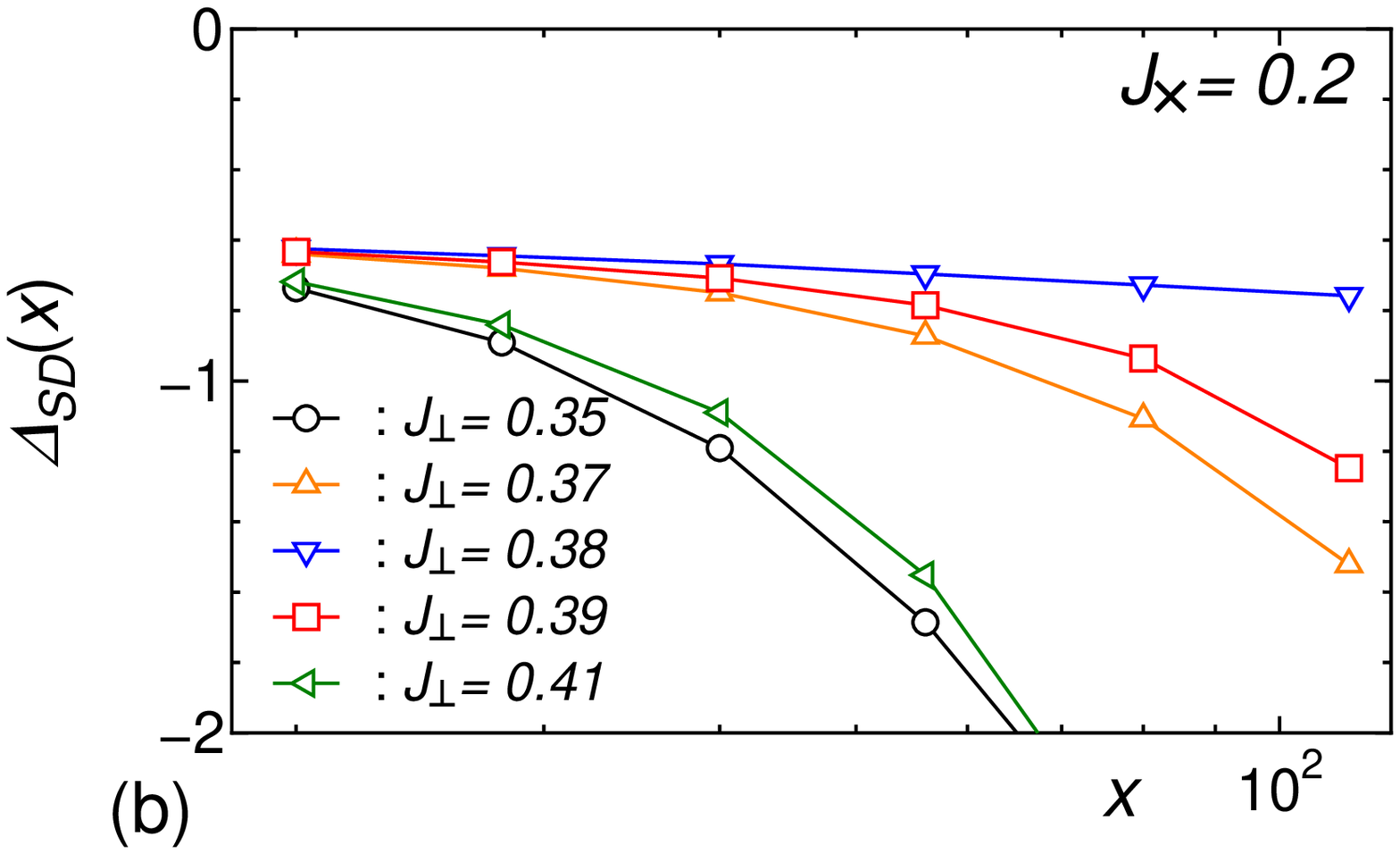}
\end{center}
\caption{
(Color online) 
Slope of the log-log plots of the dimer operators for $J = 1$ and 
$J_\times = 0.2$;
(a) $\Delta_{\rm CD}(x)$ and (b) $\Delta_{\rm SD}(x)$.
}
\label{fig:Delta-L-AF}
\end{figure}

Figures\ \ref{fig:D-L-AF} and \ref{fig:Delta-L-AF} show 
the system-size dependence of the CD and SD operators at the center of 
the open ladder, $D_{\rm CD}(L/2; L)$ and $D_{\rm SD}(L/2; L)$, 
and the slopes of their log-log plots, $\Delta_{\rm CD}(x)$ and 
$\Delta_{\rm SD}(x)$, respectively.\cite{convergence-of-slope,free-edge-spins}
[$\Delta_{\rm SD}(x)$ is defined in the same way as Eq.\ (\ref{eq:Deltacd-x}).]
We note that our data of the CD operator $D_{\rm CD}(L/2; L)$ 
for $J_\perp \le 0.37$ and $J_\perp \ge 0.39$ 
coincide with the results shown in Ref.\ \onlinecite{hung06}, 
while the data for $J_\perp = 0.38$ was not presented there.
We find in Fig.\ \ref{fig:D-L-AF} that 
both $D_{\rm CD}(L/2; L)$ and $D_{\rm SD}(L/2; L)$ decay exponentially 
with $L$ for $J_\perp \le 0.37$ (Haldane phase) 
and $J_\perp \ge 0.39$ (RS phase), 
suggesting the absence of the CD and SD orders in the parameter regions.
On the other hand, it is remarkable that for $J_\perp = 0.38$ 
the CD operator $D_{\rm CD}(L/2; L)$ bends upward in the log-log plot, 
indicating the emergence of the CD long-range order.
The tendency toward the CD ordering is elucidated 
also in Fig.\ \ref{fig:Delta-L-AF}(a), which shows that 
the slope of the $\log D_{\rm CD}(L/2; L)$-$\log L$ plot, 
$\Delta_{\rm CD}(x)$, increases with $x$.
We note that, in contrast to the CD operator, 
the SD operator $D_{\rm SD}(L/2; L)$ exhibits 
the bending-down behavior in the log-log plot even for $J_\perp = 0.38$.
The opposite trends of the CD and SD operators imply that the growth 
of the CD order observed 
at $J_\perp = 0.38$ is not a critical enhancement at a 
transition point but an indication of a true CD long-range order.
We therefore expect that the CD phase appears 
in a narrow but finite parameter region around $J_\perp = 0.38$, 
in accordance with the RG prediction\cite{ladder04} and the discussion in Section\ \ref{section:rg},
and in agreement with recent numerical finding in Ref.\ \onlinecite{liu08}.

\section{Discussion}
\label{section:discussion}

The main result of our study is the discovery of the columnar dimer phase in the 
frustrated ladder problem with {\em ferromagnetic} inter-chain interactions, 
see Fig.\ \ref{fig:phasediagram}. This finding, confirmed by extensive DMRG analysis
in Sec.\ \ref{subsection:dmrg}, is based on analytic RG arguments summarized in Sec.\ \ref{section:rg}.
It confirms novel mechanism of dimerization by frustrated
interchain couplings, proposed in Ref.\ \onlinecite{ladder04}.
Previous sightings of the spontaneously dimerized
state, of either columnar or staggered type, were
restricted to models with four-spin interaction terms, such as the ring-exchange model and the SU(2)$\times$SU(2) ladder.\cite{nersesyan97,kolezhuk98,kolezhuk98B,PatiSK1998,MullerVM2002,LauchliST2003,MomoiHNH2003}

The success of this study in describing ferromagnetic inter-leg exchanges gives us confidence 
in essential validity of the weak-coupling RG approach and makes it possible 
to re-visit the more complicated case of {\em antiferromagnetic} inter-leg exchanges,
as described in Sec.\ \ref{section:AF}. There we also find hints of developed CD order
at $(J, J_\times, J_\perp) = (1, 0.2, 0.38)$, in agreement with Ref.\ \onlinecite{liu08}. The extent of the CD region
is very narrow: finite-size scaling analysis in Ref.\ \onlinecite{liu08} estimates that
$0.373 \leq J_\perp \leq 0.386$ for $J_\times =0.2$. Such a limited range may explain
negative results of the two previous studies \cite{hung06,kim08}.

In addition to these numerical observations our work takes
important step forward in uncovering the reason for the more narrow
than naively expected, on the basis of the estimate \eqref{range}, range
of existence of the CD order. That feature, as we argue in Section\ \ref{section:rg},
has to do with marginally relevant character of the current-current interaction
between spin chains in the case of antiferromagnetic inter-leg exchanges.
We predict that the CD phase ceases to exist at all once inter-leg exchange $J_\times$ 
exceeds the critical value of the order $0.3$. 
Connecting this CD phase with the dimerized phases of frustrated
two-dimensional spin models (see Ref.\ \onlinecite{read-sachdev90} for the original large-N study and
Ref.\ \onlinecite{poilblanc09} for recent developments) represents an important outstanding problem.

Before concluding we would like to 
note that there exists another simple route to the dimerized phase.
It consists in turning marginally irrelevant in-chain backscattering 
$G_5$ into a marginally relevant one.\cite{vekua06}
This is achieved by introducing sufficiently strong 
antiferromagnetic coupling $J_2$ between next-nearest spins along the legs of the ladder. 
Provided that it exceeds the critical value  \cite{eggert}, $J_2 > 0.241 J$, the legs of the 
ladder will be spontaneously dimerized even in the absence of any inter-chain coupling.
The remaining weak inter-chain interactions then work to stabilize one of the two ordered dimerization patterns,
columnar or staggered, as is described in Ref.\ \onlinecite{vekua06} and observed in Ref.\ \onlinecite{liu08}.
Connecting this large-$J_2$ regime with the case studied here 
represents another interesting topic we leave for future.

\acknowledgments

It is our pleasure to acknowledge numerous stimulating discussions with
Leon Balents. We would like to thank A. Honecker, A. Furusaki, A. Nersesyan, and J. S\'olyom
for useful conversations.
This work was supported by Grants-in-Aid for Scientific Research
from the Ministry of Education,
Culture, Sports, Science and Technology (MEXT) of Japan,
Grant No.\ 21740277 (T.H.), and by the NSF Grant No.\ DMR-0808842 (O.A.S.).

\appendix*

\section{Dimer order in $J_1$-$J_2$ chain}
\label{section:zigzag}

In this Appendix, we check the behavior of the dimer operator 
in a finite open spin chain as a function of chain length.
To this end, we consider well-understood frustrated Heisenberg chain ($J_1$ - $J_2$ model),
\begin{eqnarray}
H_{\rm zig} = J_1 \sum_n {\bf S}_n \cdot {\bf S}_{n+1}
+ J_2 \sum_n {\bf S}_n \cdot {\bf S}_{n+2},
\label{eq:Hzig}
\end{eqnarray}
where ${\bf S}_n$ is the spin-1/2 operator at the $n$th site 
and $J_1$ and $J_2$ are coupling constants of the nearest- and 
next-nearest-neighbor exchange interactions, respectively.
It is well established that in the case of antiferromagnetic couplings, 
$J_1, J_2 > 0$, the $J_1$-$J_2$ chain (\ref{eq:Hzig}) exhibits 
a critical (Luttinger-liquid) phase for $J_2/J_1 < (J_2/J_1)_{\rm c} = 0.241$, 
while for $J_2/J_1 > (J_2/J_1)_{\rm c}$ the ground state is spontaneously 
dimerized.\cite{MajumdarG1969A,MajumdarG1969B,Haldane1982,JullienH1983,OkamotoN1992,eggert} 

\begin{figure}
\begin{center}
\includegraphics[width=0.36\textwidth]{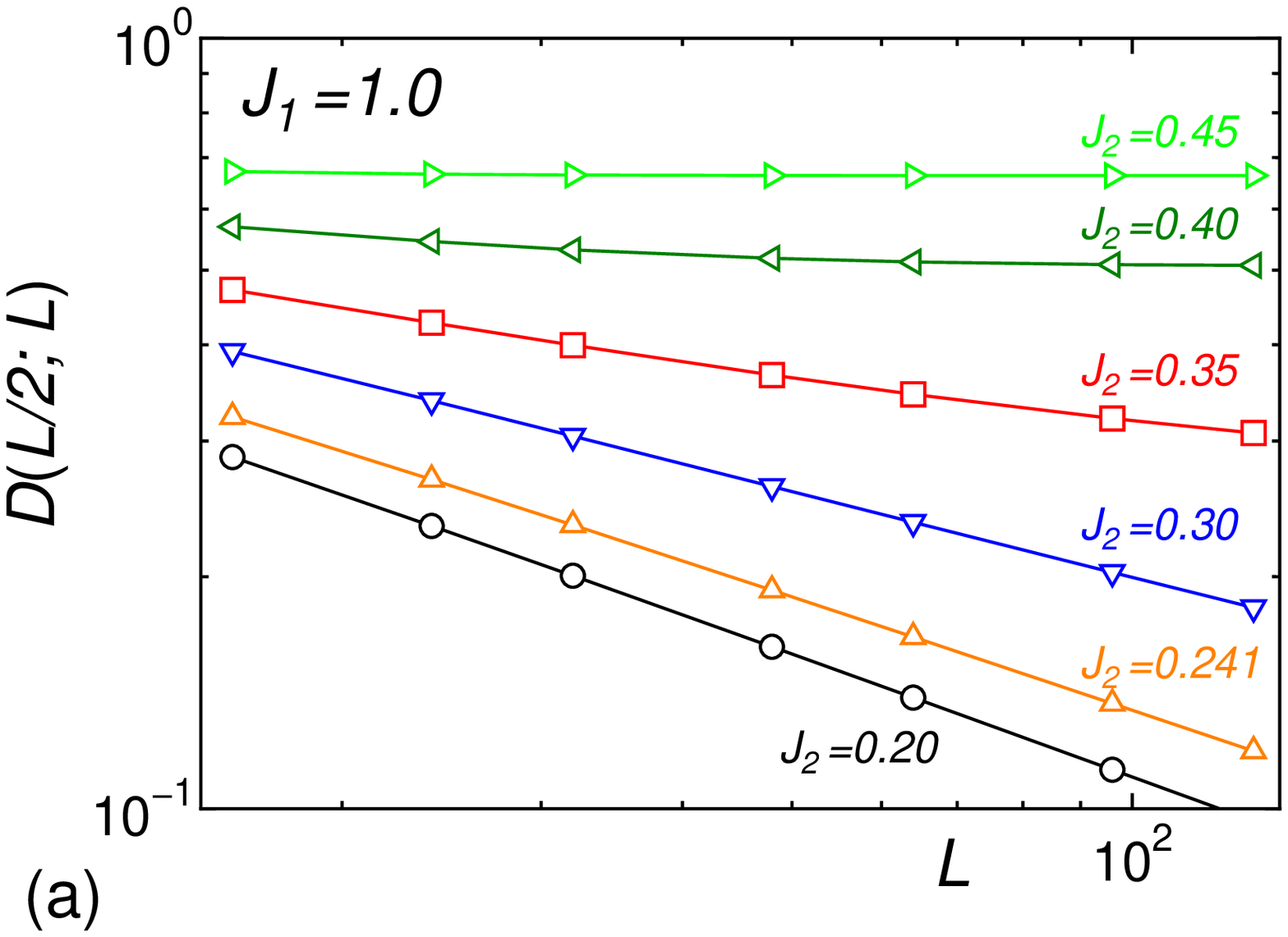}
\includegraphics[width=0.36\textwidth]{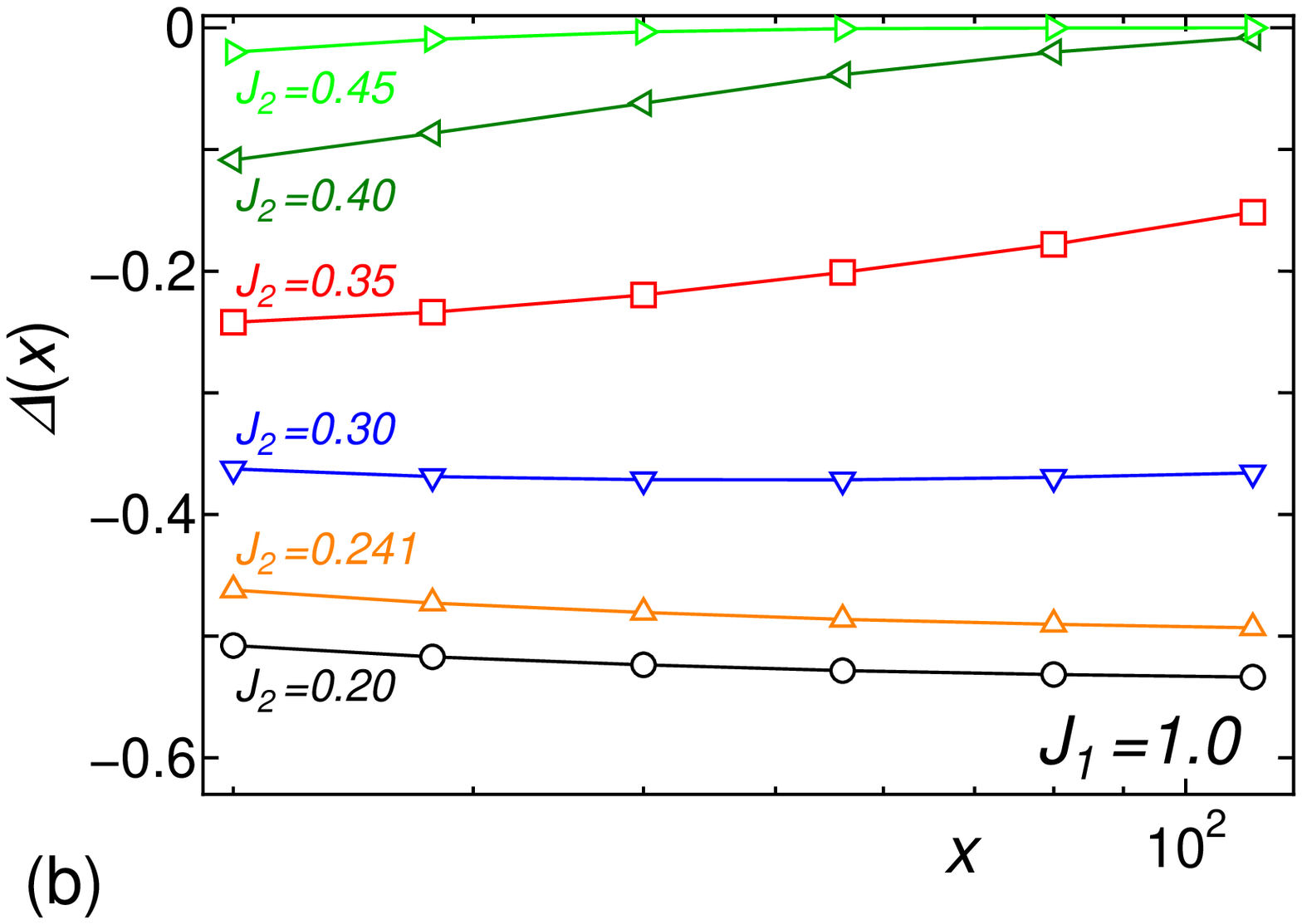}
\end{center}
\caption{
(Color online) 
DMRG data for the dimer operator in the $J_1$-$J_2$ chain with $J_1 = 1$; 
(a) dimer operator $D(L/2; L)$ at the center of the open chain, and 
(b) slope of the $\log D(L/2; L)$-$\log L$ plot, $\Delta(x)$.
The symbols represent the results for $J_2 = 0.20, 0.241, 0.30, 0.35, 0.40$, 
and $0.45$ from bottom to top.
}
\label{fig:Dzig-L}
\end{figure}

Using the DMRG method, we calculate the dimer operator, 
\begin{eqnarray}
D(n; L) = \langle {\bf S}_n \cdot {\bf S}_{n+1} 
- {\bf S}_{n+1} \cdot {\bf S}_{n+2} \rangle,
\label{eq:Dim}
\end{eqnarray}
in the chain with up to $L=128$ spins 
under the open boundary condition.
Figure\ \ref{fig:Dzig-L} shows the system-size dependence 
of the dimer operator at the center of the chain, $D(L/2; L)$, 
and its slope in the log-log plot, $\Delta(x)$, 
for several typical values of $J_2/J_1$.
[The slope $\Delta(x)$ is defined as in Eq.\ (\ref{eq:Deltacd-x}).]
For $J_2/J_1 < 0.241$, where the model is in the critical phase, 
the dimer operator $D(L/2; L)$ decays algebraically with $L$, as expected.
For $0.241 < J_2/J_1 \lesssim 0.3$, for which regime it is known that 
the system is in the dimer phase but the spin gap is exponentially small,
the dimer operator seemingly decays in a power law.
This can be understood as a consequence of the fact that
the correlation length is so large that we can not reach 
the asymptotic behavior of the dimer operator 
within the system size treated, $L \le 128$.
For $J_2/J_1 \gtrsim 0.35$, deep in the dimer phase, 
$D(L/2; L)$ shows the bending-up behavior in the log-log scale, 
and eventually, the dimer long-range order is clearly observed 
for $J_2/J_1 = 0.45$.

The results indicate that the bending-up behavior of the dimer operator 
in the log-log plot is observed only in the dimer phase 
and when the system size is comparable to or larger than the correlation length.
We can therefore safely regard the bending-up behavior 
as an evidence of the dimer ordering.


\begin{references}
\bibitem{convergence-of-slope}
In Fig.\ \ref{fig:Deltacd-L} (a) and (b) and Fig.\ \ref{fig:Delta-L-AF}, 
the data of the slope $\Delta_{\rm CD/SD}(x)$ for $x = x_7 = 160$ are not shown 
since the truncation error of the DMRG calculation, 
which is amplified by the numerical derivative 
in Eq.\ (\ref{eq:Deltacd-x}), 
is not negligible for these cases.
\bibitem{free-edge-spins} 
In Fig.\ \ref{fig:D-L-AF} (b), the SD operator $D_{\rm SD}(L/2; L)$ 
for $J_\perp = 0.39$ and $L = 192$ is not shown 
as the DMRG calculation was not stable in this case.
We note that this feature can be naturally understood within the VBS picture
as the boundary condition with additional edge spins, employed 
in the calculation of the SD operator, is not compatible with 
the RS ground state realized at this parameter point; 
In the RS ground state these additional edge spin-$1/2$ moments essentially decouple
from the bulk and spoil numerical stability of the calculation.
Similar effect is well-known in the Haldane phase where
the free edge moments are induced in the ladder 
under the simple open boundary condition.
We have observed this phenomenon in the wide range of exchange parameters
corresponding to the RS, CD, and Haldane phases.
\end{references}
\end{document}